\documentclass{aa}

\usepackage[varg]{txfonts}

\usepackage{natbib}
\bibpunct{(}{)}{;}{a}{}{,} 

\defcitealias{2022AABurnA}{Paper I}
\def\paperone{\citetalias{2022AABurnA}}
\defcitealias{Disk3}{Paper III}
\def\paperthree{\citetalias{Disk3}}

\def\tabletv{Table~3}
\def\tablets{Table~4}

\usepackage{amssymb, amsmath}
\usepackage{wasysym}

\usepackage{siunitx}
\DeclareSIUnit\erg{erg}
\DeclareSIUnit\year{yr}
\DeclareSIUnit\au{au}
\DeclareSIUnit\msol{M_\odot}
\DeclareSIUnit\mearth{M_\oplus}

\usepackage{multirow}

\usepackage{hyperref}
\hypersetup{colorlinks=true, urlcolor=blue, linkcolor=blue, citecolor=blue}
\usepackage{orcidlink}

\title{Towards a population synthesis of discs and planets}
\subtitle{II. Confronting disc models and observations at the population level\thanks{Tables 3 and 4 are only available in electronic form at the CDS via anonymous ftp to cdsarc.cds.unistra.fr (130.79.128.5) or via https://cdsarc.cds.unistra.fr/cgi-bin/qcat?J/A+A/}}

\author{Alexandre Emsenhuber\inst{\ref{lmu}}\orcidlink{0000-0002-8811-1914} \and Remo Burn\inst{\ref{mpia}}\orcidlink{0000-0002-9020-7309} \and Jesse Weder\inst{\ref{unibe}}\orcidlink{0000-0001-8270-1756} \and Kristina Monsch\inst{\ref{cfa},\ref{lmu}}\orcidlink{0000-0002-5688-6790} \and Giovanni Picogna\inst{\ref{lmu}}\orcidlink{0000-0003-3754-1639} \and Barbara Ercolano\inst{\ref{lmu},\ref{origins}}\orcidlink{0000-0001-7868-2740} \and Thomas Preibisch\inst{\ref{lmu}}\orcidlink{0000-0003-3130-7796}}

\institute{
	Universit\"ats-Sternwarte, Ludwig-Maximilians-Universit\"at M\"unchen, Scheinerstra\ss{}e 1, 81679 M\"unchen, Germany \\
	\email{emsenhuber@usm.lmu.de} \label{lmu}
	\and
	Max-Planck-Institut f\"ur Astronomie, K\"onigstuhl 17, 69117 Heidelberg, Germany\label{mpia}
	\and
	Physikalisches Institut, Universit\"at Bern, Gesellschaftsstrasse 6, 3012 Bern, Switzerland\label{unibe}
	\and
	Center for Astrophysics | Harvard \& Smithsonian, 60 Garden Street, Cambridge, MA 02138, USA\label{cfa}
	\and
	Excellence Cluster `Origins', Boltzmannstra\ss{}e 2, 85748 Garching, Germany\label{origins}
}

\titlerunning{Towards a population synthesis of discs and planets. II.}
\authorrunning{A. Emsenhuber et al.}

\date{Received 18 August 2022 / Accepted 9 December 2022}

\abstract
{}
{We want to find the distribution of initial conditions that best reproduces disc observations at the population level.}
{We first ran a parameter study using a 1D model that includes the viscous evolution of a gas disc, dust, and pebbles, coupled with an emission model to compute the millimetre flux observable with ALMA. This was used to train a machine learning surrogate model that can compute the relevant quantity for comparison with observations in seconds. This surrogate model was used to perform parameter studies and synthetic disc populations.}
{Performing a parameter study, we find that internal photoevaporation leads to a lower dependency of disc lifetime on stellar mass than external photoevaporation. This dependence should be investigated in the future. Performing population synthesis, we find that under the combined losses of internal and external photoevaporation, discs are too short lived.}
{To match observational constraints, future models of disc evolution need to include one or a combination of the following processes: infall of material to replenish the discs, shielding of the disc from internal photoevaporation due to magnetically driven disc winds, and extinction of external high-energy radiation. Nevertheless, disc properties in low-external-photoevaporation regions can be reproduced by having more massive and compact discs. Here, the optimum values of the $\alpha$ viscosity parameter lie between \num{3e-4} and \num{e-3} and with internal photoevaporation being the main mode of disc dispersal.}


\begin{document}

\maketitle

\section{Introduction}

Protoplanetary discs are the birthplace of planets (the `nebular hypothesis' of Kant and Laplace). Discs serve as a source of gas and solids from which the planets accrete. Planet--disc interactions lead to planetary migration. To model planetary formation, it is therefore essential to have disc characteristics that are as close as possible to those observed to provide the highest possible fidelity.

Disc observations are not an entirely new subject of research. Disc masses \citep[e.g.][]{1996NatureBeckwithSargent,2009ApJAndrews} and lifetimes \citep[e.g.][]{2001ApJHaisch,2009AIPCMamajek,2010AAFedele,2012ApJKraus,2014AARibas} have been observed for over two decades. However, there have been many new results concerning protoplanetary discs in the last several years, including the mass and physical extent of early discs \citep{2018ApJSTychoniec,2020AATychoniec,2020ApJTobinA} and at later times \citep{2020ApJHendler}.

Nevertheless, some aspects of disc evolution are not captured by observations, such as the process that leads to transport of material. These are usually taken to be turbulent viscosity generated by the magnetorotational instability and magnetically driven disc winds \citep{2009ApJSuzukiInutsuka,2016AASuzuki}. The strength of the turbulent viscosity has not yet been properly determined and is usually parametrised using a factor $\alpha$ \citep{1973A&AShakuraSunyaev}.

There are indirect methods to estimate the value of $\alpha$. Ultraviolet(UV)-excess measurements of the accretion luminosity were used to derive the accretion rate onto the star for the Chamaeleon~I \citep{2016AAManaraA,2017AAManara} and Lupus \citep{2014AAAlcala,2017AAAlcala} star forming regions. These measurements coupled to the disc masses for the same regions of Cha~I \citep{2016ApJPascucci} and Lupus \citep{2016ApJAnsdell} provide a relationship between mass and accretion rate \citep{2016AAManaraB}. Together, these can be used to calibrate numerical models \citep{2019AAManara} and to provide an estimate of the mass flux onto the disc \citep{2017ApJMulders,2020MNRASSellek}.
A second method to estimate $\alpha$ is to compare dust and gas emission, either using spatially resolved observations of disc substructures from ALMA \citep{2018ApJAndrewsB} such as pressure bumps \citep{2018ApJDullemond} or from the overall disc sizes \citep{2021MNRASToci}.

Mass loss does not occur only due to accretion onto the star. For instance, observations also point towards protoplanetary disc dispersal occurring from the inside out and on relatively short timescales \citep{2011MNRASErcolano,2015MNRASErcolano,2013MNRASKoepferl}. This suggests there is an additional mechanism removing gas close to the star, with one possibility being internal photoevaporation. Coupled with the findings that young stars emit a larger fraction of their flux in UV \citep[e.g.][]{2009ApSSGomezDeCastro} and X-rays \citep[e.g.][]{1996AAPreibisch,2005ApJSPreibischA,1999ARAAFeigelsonMontmerle,2003SSRvFavataMicela}, it is proposed that extreme UV \citep[EUV;][]{1994ApJHollenbach,2001MNRASClarke} and/or X-rays \citep[e.g.][]{2004MNRASAlexander,2008ApJErcolano,2009ApJErcolano} are responsible for this mass loss. Using hydrodynamical simulations, it is possible to predict the mass-loss rate as a function of disc properties and stellar luminosity \citep{2012MNRASOwen,2019MNRASPicogna,2021MNRASPicogna,2021MNRASErcolano}.

Nevertheless, irradiation from the host star is not the only mass-loss mechanism: most stars are born in clusters where many stars form concurrently. Consequently, protoplanetary discs are exposed to a larger ambient radiation field than mature stars. This leaves an additional mechanism for mass removal by external photoevaporation \citep[e.g.][]{2003ApJMatsuyama,2022EPJPWinterHaworth}. This is supported by observational findings that discs near massive stars have lower masses than others \citep{2017AJAnsdell,2020AAvanTerwisga} and that clusters with a low ambient radiation field have longer disc lifetimes \citep{2021ApJMichel}. As for external photoevaporation, hydrodynamical simulations were performed to predict mass-loss rate \citep{2018MNRASHaworth} as function of disc properties and ambient flux. Together with simulations of cluster evolution \citep[e.g.][]{2006ApJAdams,2022MNRASQiao}, this enables us to determine the mass-loss rate over an entire disc population.

All these observations and theoretical predictions put a lot of constraints on protoplanetary disc evolution, as the number of free parameters is limited. Whether or not the combination of initial disc properties and predicted accretion and mass-loss rates can be used to reproduce the distribution of, for instance, disc lifetimes remains to be determined. Previous studies in this direction usually consider only one type of  photoevaporation, either internal \citep{2009ApJGortiB,2011MNRASOwen,2021ApJKunitomo} or external \citep{2020MNRASKunitomo}.

\citet[hereafter \paperone]{2022AABurnA} introduced a relatively simple 1D radial disc model that is capable of consistently evolving gas, dust, pebbles, and planetesimals. In addition, this model is capable of predicting how the modelled disc would be observed by current instrumentation, such as ALMA (see \citealp{2018ApJBirnstiel}). Further, the light computational requirements of that model make it possible to perform many such evolutions in order to study the effects of initial disc properties.

Our goals are twofold: first, we aim to determine whether we can understand the general picture of protoplanetary discs set by the observations and predictions highlighted above. Second, we want to find the combinations of disc properties that best reproduce the various observations by performing disc populations synthesis, such as \citet{2017MNRASRosotti} and \citet{2022MNRASSomigliana}. This should then serve as initial conditions for future planetary population syntheses, such as in \citet{2009A&AMordasinia} or \citet{2021AAEmsenhuberB}.

To fit the best parameters, many calculations need to be performed, each involving the evolution of a population of protoplanetary discs. To alleviate the computational requirements of this procedure, we use machine learning to fit neural networks that can reproduce the result of the underlying model with limited resources \citep{2019ApJCambioni,2021PSJCambioni,2020ApJEmsenhuberA}. This `surrogate model' can then be used as the forward model in the fitting procedure \citep{2019IcarusCambioni}.

In this work, we aim to find initial conditions for the disc evolution calculations that best match observations. For this purpose, we first compute two series of calculations using the model presented in \paperone{} (Sect.~\ref{sec:res-fullmodel}). These data are then used to fit several surrogate models that hold the necessary outcomes for comparison with observations (Sect.~\ref{sec:res-surrogatemodel}). Using these surrogate models, we study the effect of the photoevaporation prescriptions (Sect.~\ref{sec:res-photo}) and find initial conditions that best match the observational constraints discussed in Sect.~\ref{sec:obscons} as a whole (Sect.~\ref{sec:pop}). A study dedicated to this last aspect using a Bayesian approach instead will be presented in \citet[hereafter \paperthree]{Disk3}.

\section{Methods}

The disc evolution model is based on the \textit{Bern} global model of planetary formation and evolution \citep[e.g.][]{2004A&AAlibert,2005A&AAlibert,2009A&AMordasinia,2013A&AFortier,2020AAVoelkel,2021AAEmsenhuberA} where planet formation has been turned off to retain only the disc part. \paperone{} presented an updated version of the coupled gas and solids model that includes proper modelling of the disc dispersal stage. As the model was extensively described in \paperone{}, we only provide a brief overview of the physical processes included in the model.

\subsection{Gas disc}

The gas disc is modelled by an azimuthally averaged 1D radial structure. Its evolution is obtained by solving the advection--diffusion equation \citep{1952ZNatALust,1974NMRASLyndenBellPringle}
\begin{equation}
    \frac{\partial\Sigma_\mathrm{G}}{\partial t}=\frac{3}{r}\frac{\partial}{\partial r}\left[r^\frac{1}{2}\frac{\partial}{\partial r}\left(\nu\Sigma_\mathrm{G} r^\frac{1}{2}\right)\right]-\dot{\Sigma}_\mathrm{int}-\dot{\Sigma}_\mathrm{ext},
\end{equation}
where $\Sigma_\mathrm{G}=\int_{-\infty}^{\infty}\rho_\mathrm{G}\mathrm{d}z$ is the surface density and $\nu=\alpha c_\mathrm{s}H$ the viscosity (parametrised using the $\alpha$ prescription of \citealp{1973A&AShakuraSunyaev}), with $c_\mathrm{s}$ and $H$ being the sound speed and scale height of the disc, and $\dot{\Sigma}_\mathrm{int}$ and $\dot{\Sigma}_\mathrm{ext}$  the sink terms due to internal and external photoevaporation, respectively. To compute the vertical structure of the disc (and with this $\rho_\mathrm{G}$, $H$, and $c_\mathrm{s}$), we proceed as in \paperone{} and use the vertically integrated approach of \citet{1994ApJNakamoto}, including stellar irradiation \citep{2005A&AHueso} from an evolving stellar luminosity computed from \citet{2015AABaraffe}.

\subsubsection{Internal photoevaporation}
\label{sec:method-int-phew}

Internal photoevaporation is modelled assuming X-ray-driven mass loss. This prescription requires one parameter that is not obtained from elsewhere in the model, the stellar X-ray luminosity $L_\mathrm{X}$. This luminosity is converted into a total mass-loss rate $\dot{M}_\mathrm{X}$, and then into a profile $\dot{\Sigma}_\mathrm{int}$ using fits to hydrodynamical simulations performed by \citet{2019MNRASPicogna}, \citet{2021MNRASErcolano}, and \citet{2021MNRASPicogna}, as described in \paperone{}.

\subsubsection{External photoevaporation}
\label{sec:method-ext-phew}

The mass-loss rate due to external photoevaporation is obtained from the \texttt{FRIED} grid \citep{2018MNRASHaworth}. Interpolation in the grid requires the stellar mass $M_\star$, the current disc mass $M_\mathrm{G}$, its outer radius $r_\mathrm{out}$, and the ambient far-UV (FUV) field $F$. All but the latter parameter can be computed consistently from the disc structure. The grid spans values of the ambient field $F$ between \num{10} and \SI{e4}{G_0}, where $G_0=\SI{1.6e-3}{\erg\per\second\per\square\centi\meter}$ approximately represents the interstellar value \citep{1968BANHabing}. The total mass-loss rate is converted into a profile $\dot{\Sigma}_\mathrm{ext}$ assuming mass is lost in the outermost \SI{10}{\percent} where the gas disc is present at a given time (\paperone).

The \texttt{FRIED} grid, in its current state, presents two shortcomings that we adapt here: (1) the lack of data for ambient fluxes below \SI{10}{G_0} and (2) a floor evaporation rate of \SI{e-10}{\msol\per\year}. Both items lead to a significant external photoevaporation rate under any circumstances, which makes it very difficult to disentangle the effects of external photoevaporation from the rest (including internal photoevaporation). To remedy these problems, we make one addition and one change to the \texttt{FRIED} prescription. The change is to take the lower boundary of the external photoevaporation rate down to \SI{1e-15}{\msol\per\year}, which represents a negligible mass-loss rate. For this, we remove the floor value of \SI{e-10}{\msol\per\year} from the value returned from the interpolation in the grid and ensure that the resulting value is at least \SI{1e-15}{\msol\per\year}. The change we make is to extend the domain down to \SI{1}{G_0} to be able to study low-ambient-field cases. In the region below \SI{10}{G_0}, we perform a linear interpolation between the value returned from the grid at that boundary and a fixed value of \SI{1e-15}{\msol\per\year} at \SI{1}{G_0}.

\subsection{Solids disc model}

The solid component of the disc is modelled using the two-population model of \citet{2012A&ABirnstiel}. This approximates the full size distribution using only its two extremes: the smaller $a_0$ well coupled to the gas (which can be seen as dust) and the larger, rapidly drifting $a_1$  (which can be seen as pebbles). The smaller size $a_0=\SI{0.1}{\micro\meter}$ is fixed while the larger size is constrained by various limits. The fragmentation limit is given by
\begin{equation}
	a_1 = f_\mathrm{F} \frac{2}{3\pi} \frac{\Sigma_\mathrm{G}}{\rho_\mathrm{s}\alpha} \frac{v_\mathrm{frag}^2}{c_\mathrm{s}^2},
\end{equation}
where $f_\mathrm{F}=0.37$ is a factor fitted to hydrodynamical simulations of \citet{2010AABirnstiel} for the typical size of pebbles, $\rho_\mathrm{s}=\SI{1.675}{\gram\per\cubic\centi\meter}$ is the bulk density, and $v_\mathrm{frag}$ the fragmentation velocity. In the drift limit, the large size is given by
\begin{equation}
a_1 = f_\mathrm{D} \frac{2 \Sigma_\mathrm{dust} v_\mathrm{K}^2}{\pi \rho_\mathrm{s} c_\mathrm{s}^2 \zeta} \big|\frac{\partial \ln P}{\partial \ln r}\big|^{-1},
\end{equation}
where $v_\mathrm{K}$ is Keplerian velocity, $\Sigma_0$ the surface density of dust only, and $\zeta$ is an efficiency parameter of the drift (to account for the fact that drift is more limited in discs with features such gaps created by planets; e.g. \citealp{2022AAZormpas}, while we only study smooth discs).

The surface density of solids $\Sigma_\mathrm{D}$ is divided into the two components with $\Sigma_0=\Sigma_\mathrm{D}\left(1-f_\mathrm{m}\right)$ and $\Sigma_1=\Sigma_\mathrm{D}f_\mathrm{m}$. The factor $f_\mathrm{m}=0.75$ when growth is fragmentation-limited and $f_\mathrm{m}=0.97$ when drift-limited. Gas drag onto both dust and pebbles is assumed to be in the Epstein regime. The radial velocity of solids is made of two components, coupling to the radial gas flow and headwind (\citealp{1986IcarusNakagawa}; \paperone{}),
\begin{equation}
    u_{0/1} = \frac{u_\mathrm{G,red}}{1+\mathrm{St}_{0/1}^2} - \frac{2u_\mathrm{dr}}{\mathrm{St}_{0/1} + (\mathrm{St}_{0/1} \varrho^2)^{-1}},
\end{equation}
where $u_\mathrm{G,red}$ is the reduced radial gas velocity according to \citet{2020AAGarate} and \paperone, $u_\mathrm{dr}=- \frac{r}{2v_\mathrm{K} \rho_{G}} \zeta \frac{\partial P}{\partial r}$, $\mathrm{St}$ is the Stokes number, $\varrho=\rho_\mathrm{G}/(\rho_\mathrm{G}+\rho_\mathrm{D})$, and $\rho_\mathrm{D}$ is the midplane dust density. Here, we introduce a drift efficiency $\zeta$ to parametrise mechanisms that reduce the headwind-induced drift velocity of dust. In particular, it is possible to use this approach  to represent the effect that radial substructures have on the drift of solids without modelling them in full detail. The mass-averaged radial velocity is then given by $\bar{u}=\left(1-f_\mathrm{m}\right)u_0+f_\mathrm{m}u_1$. As in the gas disc, time evolution is provided by an advection--diffusion equation,
\begin{equation}
    \frac{\partial \Sigma_\mathrm{D}}{\partial t}=\frac{1}{r}\frac{\partial}{\partial r}\left[r\left(\Sigma_\mathrm{D}\bar{u}-D_\mathrm{G}\Sigma_\mathrm{G}\frac{\partial}{\partial r}\left(\frac{\Sigma_\mathrm{D}}{\Sigma_\mathrm{G}}\right)\right)\right] - \dot{\Sigma}_\mathrm{photo} - \dot{\Sigma}_\mathrm{rad} - \dot{\Sigma}_\mathrm{pts},
\end{equation}
where $D_\mathrm{G}$ is the gas diffusion coefficient.

The terms $\dot{\Sigma}_\mathrm{photo}$ and $\dot{\Sigma}_\mathrm{rad}$ are sink terms due to dust being entrained by photoevaporative winds \citep[e.g.][]{2016MNRASFacchini,2020AAFranz} and ejected due to radiation pressure, respectively; they are both described in \paperone{}. In contrast to \paperone{}, we allow planetesimals to form. This is parametrised using the term
\begin{equation}
    \dot{\Sigma}_\mathrm{pts}=\frac{\varepsilon}{d}\frac{\dot{M}_\mathrm{D}}{2 \pi r}=\frac{\varepsilon}{d}|\bar{u}_\mathrm{dr}|\Sigma_\mathrm{D},
\end{equation}
which follows the prescription of \citet{2019ApJLenz} as implemented and described in detail in \citet{2020AAVoelkel}. Here $\varepsilon$ is a parameter that specifies the conversion efficiency into planetesimals over a length scale of $d=5H$ \citep{2013ApJDittrich}, $\bar{u}_\mathrm{dr}$ is the drift component of the mass-averaged radial velocity, and $\dot{M}_\mathrm{D}$ the relative mass flux of dust and pebbles through the gas.

\subsection{Conversion into observed disc masses}
\label{sec:methods-obs}

For consistency with disc mass observations, millimetre (mm) emission from dust and pebbles is computed from the disc surface density and temperature profiles. The method is similar to that of \citet{2018ApJBirnstiel} and will be discussed in more detail in \paperthree. The calculation is performed for a wavelength of $\lambda=\SI{0.89}{\milli\meter}$ to reproduce ALMA observations. The flux is converted back into a mass using a simple prescription assuming $T=\SI{20}{\kelvin}$ and the corresponding opacity.

For comparison, we also provide the unbiased  disc masses of gas and solids. To be presented alongside  disc gas masses, solid masses are multiplied by a factor \num{100}, which is typically used as a gas-to-dust ratio in this context.

\subsection{Model parameters and initial conditions}

The evolution model requires several initial conditions and parameters for evolution. These are: the mass of the central star $M_\star$; the mass of the gas disc $M_\mathrm{G}$; the initial dust-to-gas ratio $f_\mathrm{D/G}$; the power-law index for the initial profile $\beta$; the inner edge of the disc $r_\mathrm{in}$; the characteristic radius of the disc $r_1$; the turbulent viscosity parameter $\alpha$; the planetesimals formation efficiency $\varepsilon$; the fragmentation velocity $v_\mathrm{frag}$; the efficiency of drift $\zeta$; the stellar X-ray luminosity $L_\mathrm{X}$; and the ambient UV field strength $F$.

The initial surface density profile of the gas disc is set as \citep{2004MNRASVerasArmitage}
\begin{equation}
\Sigma_\mathrm{G}(t=0) = \Sigma_\mathrm{ini}\left(\frac{r}{r_0}\right)^{-\beta}\exp{\left(-\left(\frac{r}{r_1}\right)^{2-\beta}\right)}\left(1-\sqrt{\frac{r_\mathrm{in}}{r}}\right),
\end{equation}
where $\Sigma_\mathrm{ini}$ is the surface density at $r_0=\SI{5.2}{au}$, the reference distance. The conversion between that and the total mass is obtained with
\begin{equation}
M_\mathrm{G} = \frac{2\pi\Sigma_\mathrm{ini}}{2-\beta}\left(r_0\right)^{\beta}\left(r_1\right)^{2-\beta}.
\end{equation}
The initial solid profile of the disc $\Sigma_\mathrm{D}(t=0)$ is set by multiplying the initial gas profile $\Sigma_\mathrm{G}(t=0)$ by the dust-to-gas ratio $f_\mathrm{D/G}$.

In the remainder of this work, we do not provide all parameters as such. For instance, the inner edge is parametrised by its period $P_\mathrm{in}$, which we convert into distance by means of Kepler's third law. Also, we generally set the initial disc mass by its solid content $M_\mathrm{D}$. The ratio between the initial solids and gas masses is readily given by the dust-to-gas ratio, such that $f_\mathrm{D/G}=M_\mathrm{D}/M_\mathrm{G}$.

\subsection{Simulation list}

To generate the list of the simulations to be performed, we selected the Latin hypercube sampling (LHS) method \citep[e.g.][]{1979TechMcKay}. By dividing each dimension into $n$ intervals and then selecting one random sample from each interval, LHS ensures that the entire range of possible values for each parameter is sampled with a uniform probability. Additional criteria are required to avoid correlation between selected values of different parameters (to disentangle their effects) and to ensure that the entire space is well sampled (to avoid locations with no results). To build the grid, we use the \texttt{pyDOE2} Python package with the \texttt{minmax} setting. Each generated grid contains values in the [0,1] range with uniform probability.

\begin{table}
    \centering
    \caption{Parameter range for the main simulation grids.}
    \begin{tabular}{lccc}
        \hline
        Variable & Sampling & Min. & Max. \\
        \hline
        \hline
        $M_\star/\si{\msol}$ & linear & \num{0.1} & \num{1.4} \\
        $M_\mathrm{G}/M_\star$ & logarithmic & \num{e-3} & \num{e-0.5} \\
        $\beta$ & linear & \num{0.8} & \num{1.2} \\
        $P_\mathrm{in}/\si{\day}$ & logarithmic & \num{e-0.15} & \num{3e1} \\
        $r_1/\si{au}$ & logarithmic & \num{3e0} & \num{3e2} \\
        $\alpha$ & logarithmic & \num{e-5} & \num{e-2} \\
        $f_\mathrm{D/G}$ & logarithmic & \num{e-2.5} & \num{e-1.3} \\
        $v_\mathrm{frag}/\si{\centi\meter\per\second}$ & logarithmic & \num{2e1} & \num{2e3} \\
        $\varepsilon$ & logarithmic & \num{e-3} & \num{e-1} \\
        $\zeta$ & logarithmic & \num{e-2} & \num{1} \\
        $L_\mathrm{X}/\SI{e30}{\erg\per\second}$ & logarithmic & \num{e-2} & \num{e2} \\
        $F/G_0$ & logarithmic & \num{e0} & \num{e4} \\
        \hline
    \end{tabular}
    \label{tab:grid-range}
\end{table}

These values have to be mapped into the range to be studied. For our main grids, we outline these in Table~\ref{tab:grid-range}. The selection was made to encompass the needs of this and future works, as well as the limitations of the model. For instance, the stellar mass $M_\star$ is taken in steps of \SI{0.1}{\msol} to lie on the stellar evolution tracks of \citet{2015AABaraffe}, and the limits of the ambient UV field strength $F$ match those of the FRIED grid \citep{2018MNRASHaworth} with the extrapolation for low field values from \paperone. The gas mass of the disc is given in terms of the stellar mass to roughly follow the scaling of \citet{2021AABurn}. The dust-to-gas ratio was selected to span the possible stellar metallicities, with a reference stellar metallicity $f_\mathrm{D/G,\odot}=0.0149$ \citep{2003ApJLodders}. The range power-law index was selected to cover the possible values of \citet{2009ApJAndrews}. The lower boundary of the period at the inner edge $P_\mathrm{in}$ corresponds to \SI{0.7}{\day}, which is nearly the maximum value of the stellar radii in the models of \citet{2015AABaraffe}. The fragmentation velocity was chosen to encompass the previously assumed value of \SI{\sim10}{\meter\per\second} \citep[e.g.][and references therein]{2017AADrazkowskaAlibert}, and more current values of \SI{\sim1}{\meter\per\second}, as ice was not found to be more sticky than silicates in recent experiments \citep{2018MNRASGundlach,2019ApJMusiolikWurm,2019ApJSteinpilz}. The range of planetesimal-formation efficiencies was selected to be able to study low efficiencies where only a small fraction of the mass of solids is converted into planetesimals and to cover the case $\varepsilon=0.05,$ which forms a sufficient amount of planetesimals.

\subsection{Machine learning}

Surrogate models of disc evolution are obtained by means of a neural network. These neural networks are trained, validated, and tested using the \texttt{scikit-learn} Python package \citep{sklearn}. \texttt{scikit-learn} uses cross-validation to train and validate the neural network with five passes. This means that the combined training and validation set is divided into five equal-sized batches, and five successive training and validation steps are performed, each using four of the five batches for training and one batch for validation. The neural networks are fitted using either the \texttt{L-BFGS-B} algorithm \citep{1995SIAMJSCByrd}, which is part of the \texttt{SciPy} package \citep{2020NatMethScipy} or the \texttt{ADAM} method \citep{adam}.

\section{Observational constraints}
\label{sec:obscons}

To compute disc populations that are comparable to observations, we must first describe the constraints on their initial properties and their outcomes. These are then used to set the initial conditions and the comparison point for the outcomes.

\subsection{Stellar mass}
\label{sec:obs-mstar}

The stellar initial mass function (IMF) has been determined \citep[e.g.][]{2003PASPChabrier}, and so it could in principle be used to reproduce the stellar population. However, the stellar mass functions for different star-forming regions deviate from the IMF. In the case of Taurus, \citet{2000ApJLuhman} found a peak around \num{0.6} to \SI{1}{\msol}, while for the Orion Nebula Cluster (ONC), \citet{2012ApJDaRio} found that the best log-normal fit has a mean at $\log_{10}(M_\star/\si{\msol})=-0.45$ (corresponding to $M_\star=\SI{0.35}{\msol}$), using the stellar evolution model of \citet{1998AABaraffe}. Corroborating this, the sample of \citet{2021AAFlaischlen}, which is based on that of \citet{2012ApJManara}, has a stellar mass distribution peaking around \SI{0.4}{\msol}: a simple log-normal fit to that data gives $\log_{10}(\mu/\si{\msol})=-0.481$ and a narrower standard deviation of $\log_{10}(\sigma)=0.2383$. As the main aim of this work is to compare our model with disc lifetimes, their mass, and the stellar accretion rate of nearby star-forming regions, we chose to follow the stellar mass function of \citet{2012ApJDaRio}, with a mean of $\log_{10}(\mu/\si{\msol})=-0.45$ and standard deviation $\log_{10}(\sigma)=0.44$. This should offer a distribution that is representative of both nearby clusters in general and of stars for which observations of disc masses and stellar accretion rates are available.

Our model uses the stellar evolution tracks of \citet{2015AABaraffe} to obtain the luminosity for disc irradiation. These are only defined for mass increments of \SI{0.1}{\msol} from \num{0.1} to \SI{1.4}{\msol}. To properly track stellar luminosities, we restricted ourselves to stellar masses that match these values.

\subsection{Initial dust mass}
\label{sec:obs-mass}

Initial dust masses can be obtained from works targeting the youngest stars known to date, such as \citet{2018ApJSTychoniec}, \citet{2019ApJWilliams}, or \citet{2020ApJTobinA}. \citet{2021AAEmsenhuberB} fitted the masses of the Class~0 discs of \citet{2018ApJSTychoniec}, which gave $\log_{10}(\mu/\si{\msol})=-3.49$ and $\sigma=\SI{0.35}{dex}$, taking out the conversion from dust mass to gas mass using the standard factor of \num{100} that was used there. These disc masses were used for a population of stars with masses of \SI{1}{\msol} while the populations around lower-mass stars of \citet{2021AABurn} scaled the disc masses proportionally to the stellar masses.

However, a complication arises from the fact that the mass of the central body is not known for the objects observed by \citet{2018ApJSTychoniec} and \citet{2020ApJTobinA}. To properly convert the absolute masses into disc-to-star mass ratios, as we do in this work, we must assume a reference stellar mass $M_\star^\mathrm{ref}$. The method of \citet{2021AAEmsenhuberB} and \citet{2021AABurn} was equivalent to setting $M_\star^\mathrm{ref}=\SI{1}{\msol}$. We use this as our default conversion factor, although consistency with the stellar mass distribution discussed in the previous section would call for a lower value of $M_\star^\mathrm{ref}$. We explore different values of this factor later in this work.

\subsection{Sizes}
\label{sec:obs-size}

Protoplanetary disc sizes have been found to be correlated with their mass. \citet{2010ApJAndrews} found that discs in the Ophiuchus star-forming region have $M_\mathrm{D}\propto r_1^{\left(1.6\pm 0.3\right)}$ and $\beta=0.9\pm 0.2$. More recent studies, such as those of \citet{2017ApJTripathi} and \citet{2018ApJAndrewsA}, found that $M_\mathrm{D}\propto r_1^2$, while \citet{2020ApJHendler} obtained different scalings across various star-forming regions. For young and non-multiple discs, \citet{2020ApJTobinA} obtained $r_1\propto M_\mathrm{D}^{\left(0.25\pm 0.03\right)}$. Adding a normalisation from the same work, we get
\begin{equation}
    \frac{r_1}{\SI{70}{\au}} = \left(\frac{M_\mathrm{D}}{\SI{100}{\mearth}}\right)^{0.25},
    \label{eq:size}
\end{equation}
plus a residual scatter of the order of \SI{0.1}{dex}.

\subsection{Dust-to-gas ratio}
\label{sec:obs-fdg}

The ratio between the initial masses of the gas and dust discs is given by $f_\mathrm{D/G}$. We select this parameter as in \citet{2021AAEmsenhuberB}, that is, we assume it is the same as the stellar metallicity \citep{2016ApJGaspar}. Thus, we can use the relation $\frac{f_\mathrm{D/G}}{f_\mathrm{D/G,\odot}}=10^{\mathrm{[Fe/H]}}$ \citep{2001ApJMurray}, where $f_\mathrm{D/G,\odot}=0.0149$ is the primordial solar value \citep{2003ApJLodders}. The distribution of metallicity is chosen to be that of the CORALIE RV search sample \citep{2005AASantos}, which was modelled as a Gaussian with a mean of \num{-0.02} and a standard deviation of \num{0.22}. To avoid extreme values, we restrict the parameter to within $-0.6<\mathrm{[Fe/H]}<0.5$.

\subsection{Stellar X-ray luminosity}
\label{sec:obs-lx}

\begin{table}
    \centering
    \caption{Best-fit parameters for stellar X-ray luminosity.}
    \begin{tabular}{lccc}
        \hline
        Work & $a$ & $b$ & Sca. \\
        \hline
        \hline
        \citet{2005ApJSPreibischA} & $1.44\pm0.10$ & $30.37\pm0.06$ & $0.65$ \\
        \citet{2007AAGudel} & $1.52\pm0.12$ & $30.31\pm0.06$ & $0.54$ \\
        \hline
    \end{tabular}
    Parameters $a$ and $b$ are those of the fit $\log{\left(L_\mathrm{X}/\si{\erg\per\second}\right)}=a\times\log{\left(M_\star/M_\odot\right)}+b$. The `Sca.' column provides the scatter of the residuals from the fit.
    \label{tab:lx-fits}
\end{table}

A couple of surveys have been performed to determine the X-ray luminosities of young stars, the relevant results of which are provided in Table~\ref{tab:lx-fits}. One is the \textit{Chandra} Orion Ultradeep Project \citep[COUP;][]{2005ApJSGetmanA,2005ApJSPreibischA}, which covers stellar masses $M_\star$ between \num{0.5} and \SI{0.9}{M_\odot}. The survey found a stellar-mass dependency of $L_\mathrm{X}\propto M_\star^{(1.44\pm 0.10)}$. Another survey, the \textit{XMM-Newton} Extended Survey of Taurus \citep[XEST;][]{2007AAGudel}, found that $L_\mathrm{X}$ varies with stellar mass as $L_\mathrm{X}\propto M_\star^{(1.54\pm 0.12)}$, which we used to correct for the stellar-mass effect and recompute the inherent scatter. The two surveys have similar stellar mass dependence, meaning that using one or the other to set the stellar X-ray luminosities should not affect the outcomes in any significant manner. For this work, we compute $L_X$ using a log-normal distribution with the parameters selected following XEST \citep{2007AAGudel}, as the stellar mass dependence is consistent with the prescription used to compute the X-ray photoevaporation profiles in \citet{2021MNRASPicogna}.

\subsection{Ambient FUV field strength}

The external photoevaporation prescription of \citet{2018MNRASHaworth} requires the stellar mass, disc mass, outer radius, and ambient FUV field strength. The first three can be readily obtained consistently from the simulation, but the latter, $F$, needs to be specified.

Most stars are formed in stellar clusters \citep[e.g.][]{2003ARAALadaLada}, which result in high stellar densities. To retrieve the ambient FUV relevant during the lifetime of protoplanetary discs, we use the simulation of \citet{2006ApJAdams}. The authors determined that $F$ is well described by a log-normal distribution with a median close to $10^{3.25}G_0$, where $G_0=\SI{1.6e-3}{\erg\per\second\per\square\centi\meter}$ is nearly the interstellar FUV field \citep{1968BANHabing}.

\subsection{Inner edge of the gas disc}

The location of the inner edge of the gas disc is most relevant for the location of the close-in planets (such as hot Jupiters). As we are mostly interested in warm giants further away than the inner edge, this parameter is of less importance in this work. We chose this parameter in the same way as in \citet{2021AAEmsenhuberB}, that is, by assuming that the disc is truncated at the corotation radius of the star. For the distribution of stellar rotation periods, we follow the results of \citet{2017AAVenuti}. This gives a log-normal distribution with a median period of $\log_{10}(\mu/\mathrm{d})=0.67617866$ and deviation $\sigma=\SI{0.3056733}{dex}$.

For comparison, the distribution of initial rotation periods used by \citet{2021AAJohnstone} has a median of $\log_{10}(\mu/\mathrm{d})=0.5181$ and a standard deviation of $\sigma=\SI{0.3236}{dex}$. The median rotation period here is smaller here (\SI{3.3}{\day}) than the \SI{4.7}{\day} value of \citet{2017AAVenuti} but not by a large amount, while the deviations are similar. The exact choice should therefore not affect the results significantly.

\section{Results}

\subsection{Full model}
\label{sec:res-fullmodel}

To generate the training, validation, and testing data for the surrogate models, we generated two sets of simulations. The first set contains \num{100000} models that are used for the combined training and validation steps, while the second set contains \num{20000} models and is used for the testing step. The values of the first set are provided in \tabletv{} while the values of the second set are provided in \tablets{}. Both tables are available at the CDS and have the same format; they contain the following columns: columns 1 to 12 are the initial conditions in the same order and units that are given in Table~\ref{tab:grid-range}. Column 13 gives the disc lifetime according to when the mass becomes lower than $\num{e-6}M_\star$ or when the surface density is lower than \SI{1e-3}{\gram\per\square\centi\meter} inside \SI{100}{\au} (or \SI{30}{\au} for $M_\star=\SI{0.1}{\msol}$ or \SI{0.2}{\msol}) and \SI{1e-2}{\gram\per\square\centi\meter} outside that (this second criterion on the surface density is to avoid excessively long-lived discs when photoevaporation rates, particularly external ones, are low). Column 14 gives the lifetime using the minimum value of the criterion of column 13 and the observability criterion in the near-infrared (NIR) from \citet{2016MNRASKimura}. Columns 15-19 give the following outcomes at \SI{1e5}{\year}: stellar accretion rate $\log_{10}\left(\dot{M}_\star/\si{\msol\per\year}\right)$, the true gas mass $\log_{10}\left(M_\mathrm{G}/\si{\msol}\right)$, the true solids mass $\log_{10}\left(M_\mathrm{D}/\si{\msol}\right)$, the observed mass (Sect.~\ref{sec:methods-obs}) $\log_{10}\left(M_\mathrm{obs}/\si{\msol}\right)$ and the radius encompassing \SI{68}{\percent} of the flux $\log_{10}\left(r_\mathrm{68}/\si{\au}\right)$. Columns 20-24 repeat the same information, but at \SI{2e6}{\year}.

Two epochs (\SI{1e5}{\year} and \SI{2e6}{\year}) were selected to be compatible with the observations we are comparing to. The first epoch is for comparison with early discs, such as their initial masses. Its selection is a trade-off between two items: on the one hand, we would like to have the data as early as possible, while on the other hand, we need to wait until the initial dust growth has taken place. From the analysis of individual discs, we found that \SI{1e5}{\year} represents a good compromise in that sense. The second epoch is for comparison with the star-forming regions of Lupus and Cha~I. As the stars in these regions are between \num{1} and \SI{3}{\mega\year} old, we take the results at \SI{2}{\mega\year}, as in \citet{2019AAManara}.

Our results indicate that the two criteria for disc dispersal produce nearly identical results. In only about \SI{10}{\percent} of the cases, the NIR criterion predicts a lower disc lifetime than the criterion based on the mass, and the difference remains small when this occurs (we do not check for the reverse, as calculations stop when the mass criterion is reached). These results are consistent with the findings of \citet{2021ApJKunitomo}. As a consequence, hereafter we only report the disc lifetimes based on the NIR criterion of \citet{2016MNRASKimura}. Also, we stop the calculation at \SI{100}{\mega\year} in any case. This affects some long-lived discs with minimal photoevaporation and accretion. In such cases, the lifetime based on the mass criterion is not reported while that based on the NIR emission is.

\subsection{Performance of the surrogate model}
\label{sec:res-surrogatemodel}

\begin{figure*}
	\centering
	\includegraphics{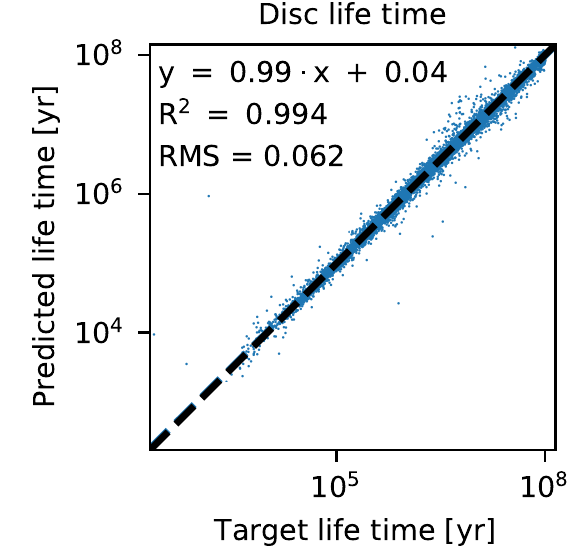}
	\includegraphics{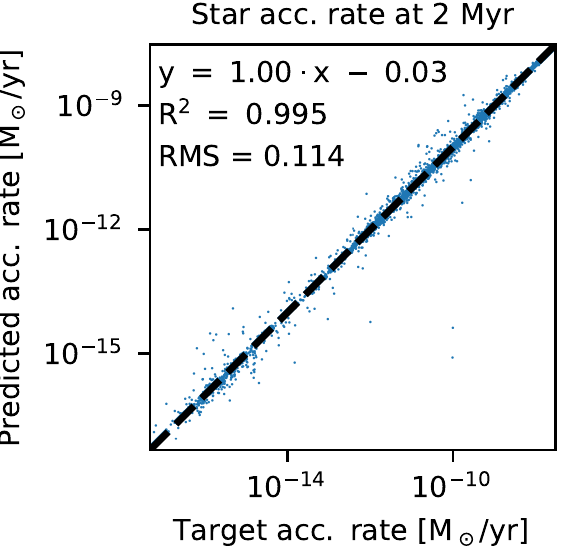}
	\includegraphics{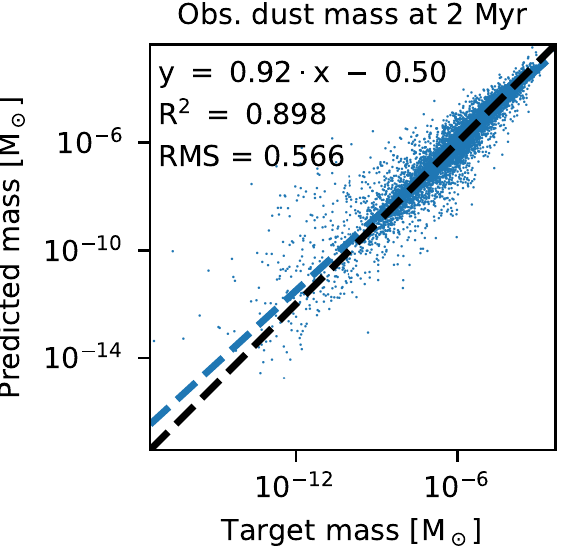}
	\caption{Performance of three surrogate models based on the comparison of the predicted and actual values of the testing set. The insert values show the best regression (ordinary least squares), the Pearson correlation coefficient $R^2$, and the RMS of the differences between each predicted and actual value.}
	\label{fig:corr-dlt-nir}
\end{figure*}

\setcounter{table}{4}
\begin{table*}
	\centering
	\caption{Hyper parameters and performance of the surrogate models.}
	\label{tab:sm}
	\begin{tabular}{cccccccccc}
        \hline
        Age & Model & Solver & Activ. & HLS & alpha & Val. $R^2$ & Test. $R^2$ & Val. MSE & Test. MSE \\
        \hline
        \hline
         & Lifetime & L-BFGS & logistic & 25, 50, 45 & \num{2.592e-3} & 0.99465 & 0.99436 & 0.00368 & 0.00389 \\
        \hline
        \multirow{3}*{\SI{100}{\kilo\year}} & Accretion & L-BFGS & tahn & 15, 30, 50 & \num{1.098e-3} & 0.99909 & 0.99829 & 0.00163 & 0.00300 \\
        & Mass & ADAM & tahn & 55, 40, 65 & \num{2.659e-3} & 0.96330 & 0.95040 & 0.05690 & 0.07677 \\
        & Radius & ADAM & tahn & 60, 55, 45 & \num{1.131e-3} & 0.89222 & 0.88045 & 0.04460 & 0.04853 \\
        \hline
        \multirow{3}*{\SI{2}{\mega\year}} & Accretion & L-BFGS & tahn & 60, 45, 70 & \num{9.407e-3} & 0.99865 & 0.99506 & 0.00349 & 0.01297 \\
        & Mass & ADAM & tahn & 65, 25, 55 & \num{1.915e-3} & 0.91565 & 0.89798 & 0.25667 & 0.32089 \\
        & Radius & ADAM & tahn & 70, 65, 35 & \num{6.901e-3} & 0.85665 & 0.80786 & 0.11881 & 0.15925 \\
        \hline
	\end{tabular}
\end{table*}

We asses the performance of the surrogate models in terms of the best regression (obtained using ordinary least squares), the Pearson correlation coefficient $R^2$, and the RMS of the differences between the predicted and target lifetimes (the square root of the mean square error). These were computed on the testing set (\tablets) that the surrogate model has not seen before. The hyper-parameters and results for all surrogate models that are part of this work are presented in Table~\ref{tab:sm}. For three of them, we also show correlation plots in Fig.~\ref{fig:corr-dlt-nir}. In all cases, the fitting procedure was performed on the logarithm (base 10) of the values, and so all the reported performances are given in these units.

Concerning the different surrogate models, the ones for the disc lifetimes and for the stellar accretion rates provide the best performance. The ones that are based on the dust disc, namely masses and radii, show a lower performance, especially at \SI{2}{\mega\year}. We note that these values are for each single prediction; they represent the level of additional uncertainty for the parameter studies (Sect.~\ref{sec:res-photo} and Appendix~\ref{sec:paramstudy}) while for the population studies (Sect.~\ref{sec:pop}) these errors can average out and result in an even better global accuracy.

The neural networks predicting the disc masses, their radii, and stellar accretion rates were fitted only on the discs that had not vanished at the time. This means that they are supported by a lower number of points than the ones predicting the lifetimes. This also implies that these surrogate models are only constrained in the region of the parameter space where lifetimes are larger than the time of the analysis. Thus, in the remainder of this work we only provide disc masses and stellar accretion rates for discs that have not yet dispersed.

\subsection{Effects of photoevaporation}
\label{sec:res-photo}

We began our investigations using the surrogate model, performing a parameter study of the effects of the photoevaporation prescriptions on disc lifetimes. For this purpose, we generated two maps, one for internal photoevaporation and one for external photoevaporation, which vary the stellar mass and the controlling parameter of each photoevaporation prescription. In each case, the value of the parameter controlling the other photoevaporation prescription was set at the minimum of the studied range in order to avoid cross effects. We assumed typical values of the remaining parameters:  disc-to-star gas mass ratio $M_\mathrm{G}/M_\star=\num{0.1}$, dust-to-gas ratio $f_\mathrm{D/G}=0.0149$, power-law index $\beta=\num{0.9}$, period at the inner edge $P_\mathrm{in}=\SI{10}{\day}$, characteristic radius $r_\mathrm{1}$ computed according to Eq.~(\ref{eq:size}), viscosity parameter $\alpha=\num{1e-3}$, fragmentation velocity $v_\mathrm{frag}=\SI{2}{\meter\per\second}$, planetesimal formation efficiency $\varepsilon=\num{1e-3}$, and drift efficiency $\zeta=1$.

\subsubsection{Internal photoevaporation}
\label{sec:res-photo-int}

\begin{figure}
	\centering
	\includegraphics{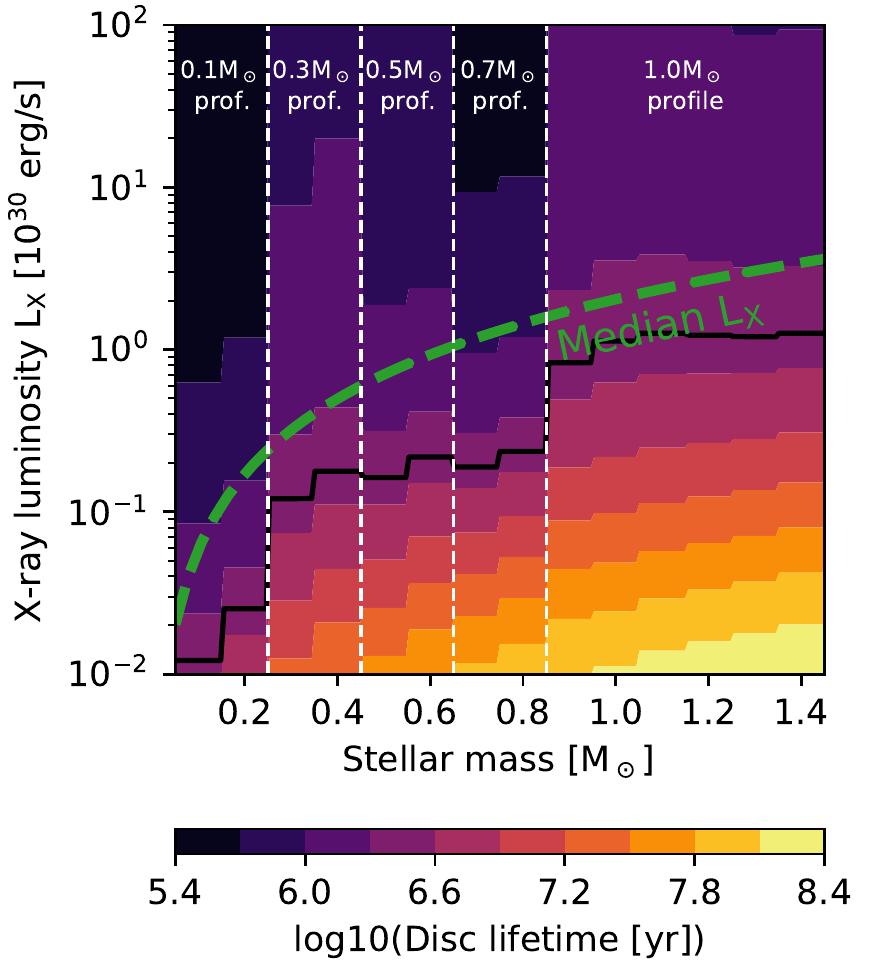}
	\caption{Map of disc lifetimes as a function of stellar mass and X-ray luminosity, which is the main driver of internal photoevaporation, according to the surrogate model described in Sect.~\ref{sec:res-surrogatemodel}. External photoevaporation was set to its minimum value ($F=\SI{1}{G_0}$). Other parameters were selected as $M_\mathrm{G}/M_\star=\num{0.1}$, $f_\mathrm{D/G}=0.0149$, $\beta=\num{0.9}$, $P_\mathrm{in}=\SI{10}{\day}$, $r_\mathrm{1}$ according to Eq.~(\ref{eq:size}), $\alpha=\num{1e-3}$, $v_\mathrm{frag}=\SI{2}{\meter\per\second}$, $\varepsilon=\num{1e-3}$, and $\zeta=1$. The green dashed line represents the dependency of $L_\mathrm{X}$ on $M_\star$ from \citet{2007AAGudel} and the solid black line shows the location of a \SI{3}{\mega\year} lifetime (a typical value). The results are discussed in Sect.~\ref{sec:res-photo-int}.}
	\label{fig:int-dlt-nir}
\end{figure}

The resulting map for internal photoevaporation is provided in Fig.~\ref{fig:int-dlt-nir}. Here we observe that the surrogate model predicts several sharp transitions of disc lifetime with stellar mass. The most evident are those between \num{0.2} and \SI{0.3}{\msol} and between \num{0.8} and \SI{0.9}{\msol} where disc lifetime increases. There are other transitions between \num{0.4} and \SI{0.5}{\msol} and between \num{0.6} and \SI{0.7}{\msol} where disc lifetime decreases, but only for large X-ray luminosities ($L_\mathrm{X}>\SI{e30}{\erg\per\second}$). These transitions match the switch from one photoevaporative profile to another, which are marked by the dashed white lines. This indicates that the profile of surface-density loss has a strong effect on disc lifetime and not only the total mass-loss rate, which gradually changes between each stellar mass. Also, the further out the location of the peak of internal photoevaporation (which is for the profiles of \SI{0.3}{\msol} and \SI{1.0}{\msol} stars; see top panel of Figure~7 of \citealp{2021MNRASPicogna}), the longer the disc lifetimes are in general. We find that this effect is due to a larger inner region where material is not evaporated at all and can only be dispersed by viscous accretion. In this case, the observed disc lifetime is set by the dispersal timescale of the inner disc, which is given by the viscous timescale at the outer radius of the inner disc.

As discussed in Sect.~\ref{sec:obs-lx}, the X-ray luminosity is correlated with stellar mass. To highlight this, we show in addition the median stellar X-ray luminosity as a function of stellar mass from \citet{2007AAGudel} with the green dashed curve. To determine the expected relationship between disc lifetime and stellar mass, one needs to follow this curve rather than a horizontal line on the plot. We see that internal photoevaporation leads to a limited change in disc lifetime with stellar mass. This is because more massive stars lead to stronger mass-loss rates (owing to a corresponding increase in stellar X-ray luminosity), which compensates for the increase in disc mass (as we assume disc mass to be proportional to stellar mass). This is shown by the black line that traces disc lifetimes of \SI{3}{\mega\year} (a typical value in observations), which is consistently lower than the median $L_\mathrm{X}$ by a factor of a few.

\subsubsection{External photoevaporation}
\label{sec:res-photo-ext}

\begin{figure}
	\centering
	\includegraphics{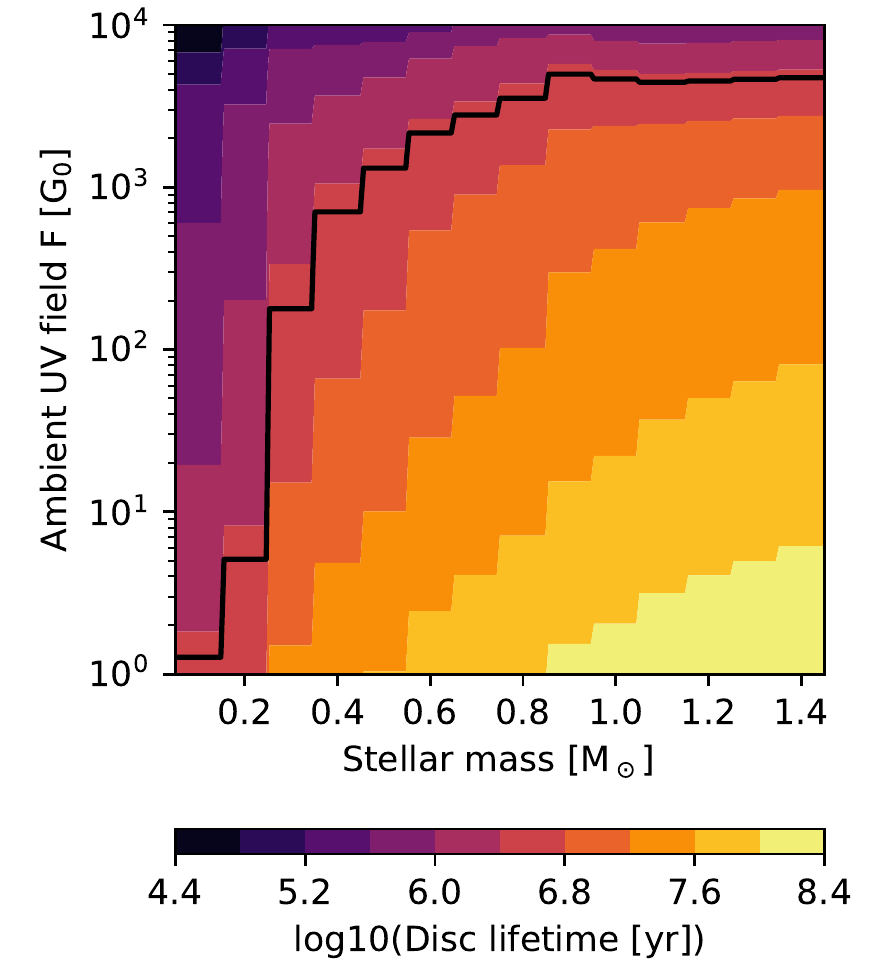}
	\caption{Map of disc lifetime as functions of stellar mass and ambient UV field strength, this latter being the main driver of external photoevaporation according to the surrogate model described in Sect.~\ref{sec:res-surrogatemodel}. Internal photoevaporation is set to its minimum value ($L_\mathrm{X}=\SI{1e28}{\erg\per\second}$). Other parameters were selected as $M_\mathrm{G}/M_\star=\num{0.1}$, $f_\mathrm{D/G}=0.0149$, $\beta=\num{0.9}$, $P_\mathrm{in}=\SI{10}{\day}$, $r_\mathrm{1}$ according to Eq.~(\ref{eq:size}), $\alpha=\num{1e-3}$, $v_\mathrm{frag}=\SI{2}{\meter\per\second}$, $\varepsilon=\num{1e-3}$, and $\zeta=1$. The solid black line shows the location of a \SI{3}{\mega\year} lifetime (a typical value). The results are discussed in Sect.~\ref{sec:res-photo-ext}.}
	\label{fig:ext-dlt-nir}
\end{figure}

The resulting map for external photoevaporation is shown in Fig.~\ref{fig:ext-dlt-nir}. Unlike internal photoevaporation, the prescription for external photoevaporation provides for gradual changes of lifetime with stellar mass. However, these changes lead to a larger dependency of disc lifetime on stellar mass than what is expected from internal photoevaporation. This is illustrated by the black line, which tracks a \SI{3}{\mega\year} lifetime, as in the map for internal photoevaporation. Its position in terms of ambient UV field strength varies across the entire parameter range studied here, from less that \SI{2}{G_0} for $M_\star=\SI{0.1}{\msol}$ to about \SI{4e3}{G_0} for $M_\star=\SI{1}{\msol}$; it becomes independent of stellar mass for $M_\star>\SI{1}{\msol}$ and fluxes above \SI{\sim e3}{G_0}.

While the trend of reduced disc lifetimes in regions with strong ambient UV fields \citep[e.g.][]{2021ApJMichel} is reproduced, the general behaviour of correlated disc lifetimes with stellar mass for a given ambient UV field strength is problematic for several reasons. First, this general behaviour is inconsistent with observations that suggest disc lifetimes are independent of, or slightly decreasing with, increasing stellar mass \citep{2006ApJCarpenter,2009ApJKennedyKenyon,2012AABayo,2015AARibas}. There are several possibilities to remedy this, although they are unlikely. To obtain a behaviour similar to observations, the mass loss would need to be correlated with stellar mass \citep{2021ApJKomaki}, which in turn would require that the ambient field be correlated with stellar mass. However, the ambient field is usually dominated by the few most massive stars \citep{2006ApJAdams}, which means that it depends more on the cluster as a whole than on the mass of the star in question. Another avenue is that disc masses scale to a lesser extent with stellar mass than assumed here. However, this would not yield the expected behaviour of stellar accretion rates with stellar masses \citep[e.g.][]{2016ARAAHartmann,2017AAAlcala,2021AAFlaischlen}. We find that the \texttt{FRIED} grid prescription that we use in this work produces incompatible results that show at most a dependence of the disc lifetime on stellar mass. The second concern is that further lifetime analyses will be strongly affected by the selection of the stellar masses, in contrast to internal photoevaporation where this dependency is weaker.

\subsection{Parameter sensitivity}

The sensitivity of disc lifetimes, disc masses, and stellar accretion rates at \SI{2}{\mega\year} is studied in detail in Appendix~\ref{sec:paramstudy}. These results can be summarised as follows: all the outcomes are insensitive to the power-law index $\beta$ and the inner edge of the gas disc $r_\mathrm{in}$. The characteristic radius $r_1$ and viscosity $\alpha$ control the viscous timescale of the disc, and therefore the stellar accretion rate. Disc lifetimes and observed dust masses are more strongly affected by the viscosity $\alpha$ than by the characteristic radius $r_1$.

The \textit{twopop} model parameters only affect the observed dust masses. Observed dust masses are less affected by the dust-to-gas ratio $f_\mathrm{D/G}$ than by the initial  mass of the gas disc $M_\mathrm{G}$, except for discs close to dispersal. The fragmentation velocity $v_\mathrm{frag}$ and the drift efficiency strongly affect the observed dust masses, but only for $v_\mathrm{frag}\gtrsim\SI{200}{\centi\meter\per\second}$, while the planetesimal formation efficiency $\varepsilon$ only has a limited effect for values close to the maximum we study, namely of \num{0.1}.

These results narrow down the parameter space that we explore in the remainder of this work. First, we keep the values of the power-law index $\beta$ and the inner edge of the gas disc $r_\mathrm{in}$ as described in Sect.~\ref{sec:obscons}, because they are of negligible importance. We then only use the initial mass of the gas disc  $M_\mathrm{G}$ to control disc masses, not the dust-to-gas ratio $f_\mathrm{D/G}$; as the latter is in most cases of lower importance and well constrained by observations. Also, we keep the planetesimal formation efficiency to the minimum value of $\varepsilon=\num{1e-3}$, the fragmentation velocity to $v_\mathrm{frag}=\SI{200}{\centi\meter\per\second}$, and the drift efficiency to $\zeta=1$.

\section{Disc populations}
\label{sec:pop}

We now compare synthetic disc populations with observations. For this, we proceed as follows: we draw \num{10000} random discs whose initial conditions follow given distributions. The outcomes of each disc are obtained by means of the different surrogate models. For the analysis, we first compare the cumulative distribution of disc lifetimes so that it can be compared to the fraction of stars that have a protoplanetary disc for stellar clusters with different ages, as in \citet{2001ApJHaisch}. The second analysis is to compare observed disc masses and stellar accretion rates with the data of \citet{2019AAManara}. Here, we use the data at \SI{2}{\mega\year}. Further, we only use discs whose lifetime, as determined by the surrogate model from the previous analysis, is larger than the time of analysis in order to avoid being in the region where the surrogate model is not supported by any underlying data.

\subsection{Canonical}
\label{sec:pop-canonical}

\begin{table}
    \centering
    \caption{Random distributions for the canonical population}
    \begin{tabular}{ll}
        \hline
        Variable & Distribution \\
        \hline
        \hline
        $\log_{10}(M_\star/\si{\msol})$ & $\mathcal{N}(-0.45, 0.44^2)$ \\
        $M_\mathrm{G}$ & $M_\mathrm{D}/f_\mathrm{D/G}$ \\
        $\beta$ & $0.9$ \\
        $\log_{10}(P_\mathrm{in}/\si{\day})$ & $\mathcal{N}(0.67617866, 0.3056733^2)$ \\
        $r_1/\si{au}$ & $70(M_\mathrm{D}/\SI{100}{\mearth})^{0.25}\times10^{\mathcal{N}(0, 0.1^2)}$ \\
        $\log_{10}(\alpha)$ & $\mathcal{U}(-3.5, -3.0)$ \\
        $\log_{10}(f_\mathrm{D/G})$ & $\mathcal{N}(-1.85, 0.22^2)$ \\
        $\log_{10}(M_\mathrm{D}/M_\star)$ & $\mathcal{N}(-3.49, 0.35^2)$ \\
        $v_\mathrm{frag}/\si{\centi\meter\per\second}$ & \num{200} \\
        $\varepsilon$ & \num{e-3} \\
        $\zeta$ & \num{1} \\
        $L_\mathrm{X}/\SI{e30}{\erg\per\second}$ & $10^{\mathcal{N}(0.31, 0.54^2)}\times(M_\star/\SI{1}{\msol})^{1.52}$ \\
        $\log_{10}(F/G_0)$ & $\mathcal{N}(3.25, 0.93^2)$ \\
        \hline
    \end{tabular}
    \label{tab:canon-vars}
\end{table}

\begin{figure}
	\centering
	\includegraphics{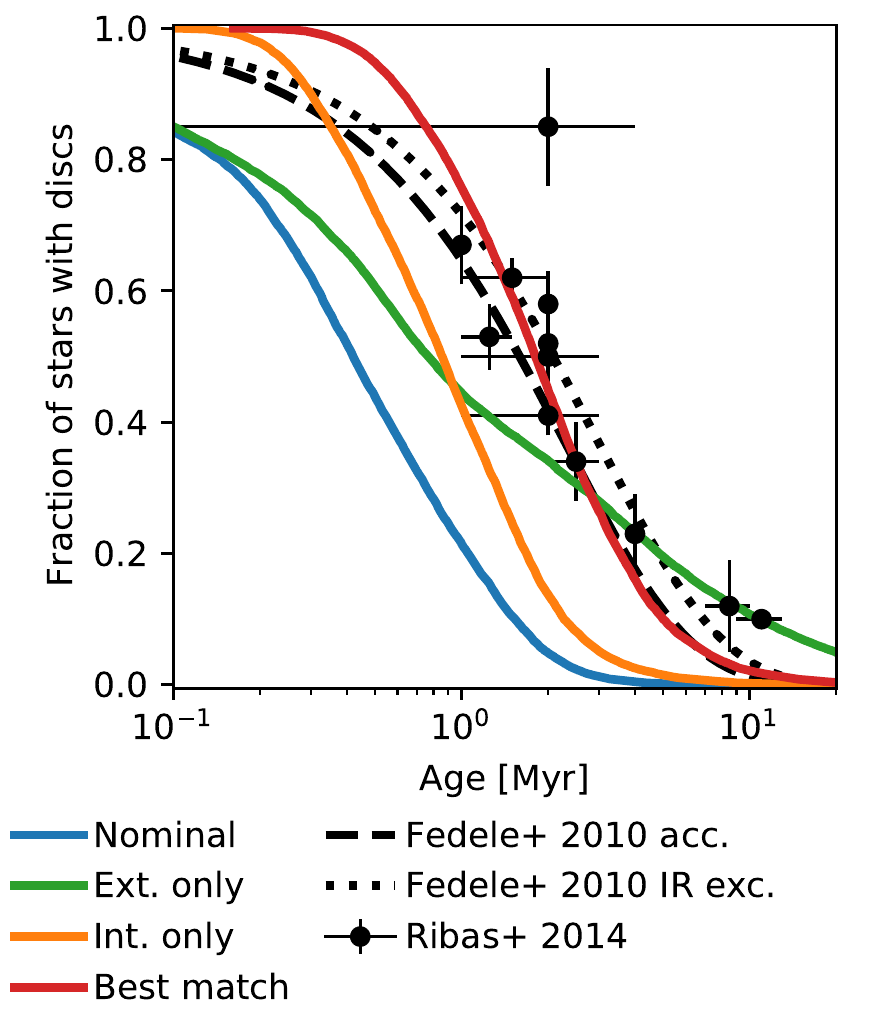}
	\caption{Cumulative distribution of disc lifetimes for a population with canonical parameter distribution (see text). Two exponential decays following \citet{2010AAFedele} with a characteristic time of \SI{2.3}{\mega\year} (accretion) and \SI{3}{\mega\year} (infrared excess) and the results of \citet{2014AARibas} are shown as well.}
	\label{fig:dlt-pop-canon}
\end{figure}

\begin{figure*}
	\centering
	\includegraphics{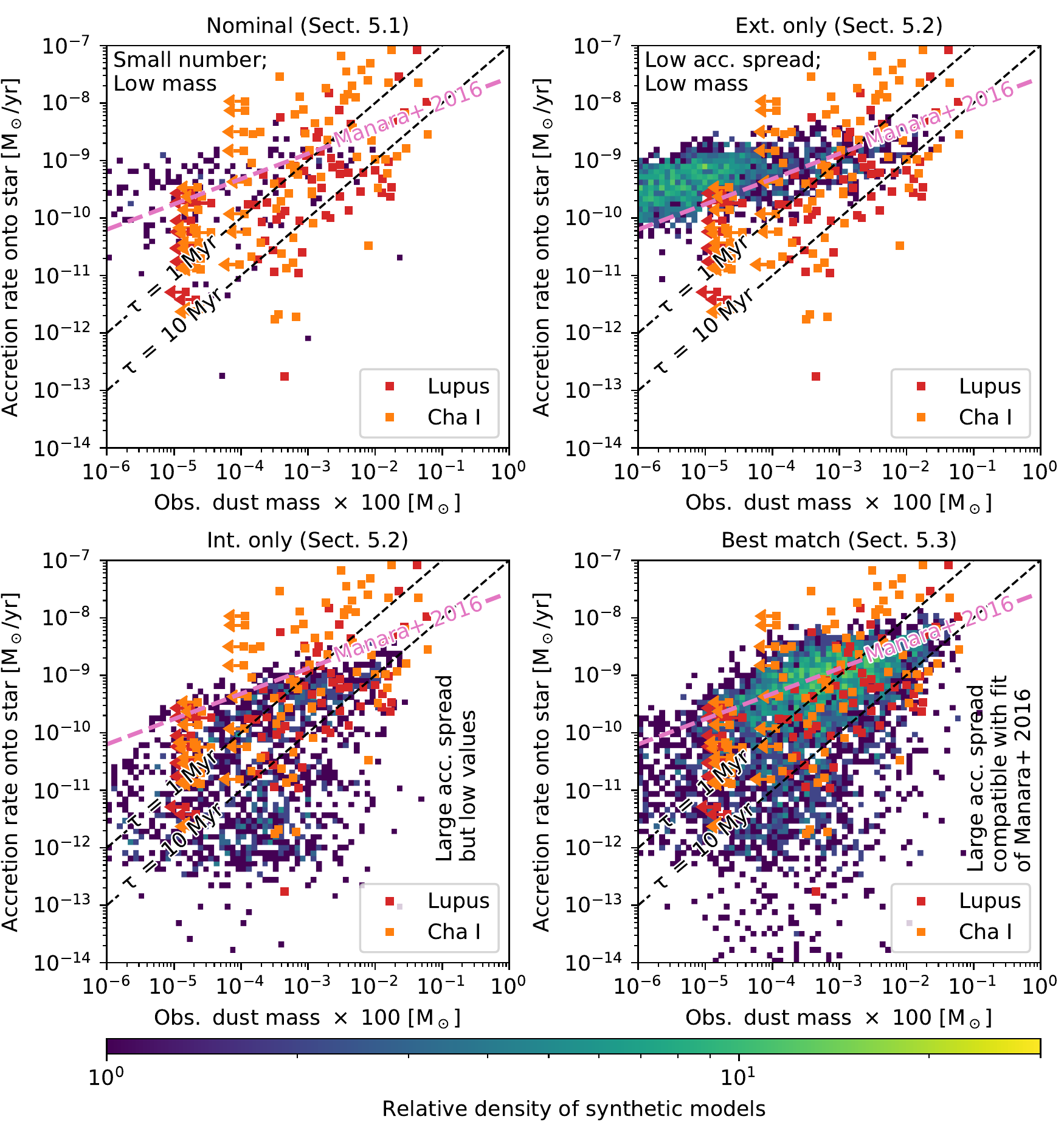}
	\caption{Histogram for stellar accretion rate vs. disc mass at \SI{2}{\mega\year} and the same synthetic disc populations shown in Fig.~\ref{fig:dlt-pop-canon}. The observed data from the Lupus (in red) and Chamaeleon~I (orange) star forming regions from \citet{2019AAManara} are shown for comparison. Two disc dispersal timescales $\tau=\dot{M}_\mathrm{G}/M_\mathrm{G}$ of \SI{1}{\mega\year} and \SI{10}{\mega\year} (assuming that the gas disc mass $M_\mathrm{G}$ is \num{100} times the observed dust mass) and the best fit to the data from \citet{2016AAManaraB} are shown as well.}
	\label{fig:mar-pop-canon}
\end{figure*}

To determine if all the processes that are predicted from theory are able to reproduce disc observations, we compute a population of discs whose properties are as close as possible to observations from early discs. The only parameter that has some freedom is $\alpha$. Here, we selected to draw $\log_\mathrm{10}{(\alpha)}$ with a uniform probability of between $-3.5$ and $-3$. This was decided as a compromise between disc lifetime and stellar accretion rate, as we discuss below. For the other parameters, their distributions were selected as described in the discussion of Sect.~\ref{sec:obscons}; for convenience, these are summarised in Table~\ref{tab:canon-vars}.

The resulting  distribution of disc lifetimes is shown with the blue curve in Fig.~\ref{fig:dlt-pop-canon}. It becomes immediately apparent that the synthetic lifetimes are too short  overall in comparison with observed discs. The median lifetime of the synthetic discs is \SI{0.42}{\mega\year}. Disc lifetime depends on the assumed distribution of $\alpha$, which we chose such that it results in the largest stellar accretion rates at \SI{2}{\mega\year}. A histogram of stellar accretion rate versus observed dust mass for the observed discs in the Lupus and Chamaeleon~I star-forming regions  is shown in the top-left panel of Fig.~\ref{fig:mar-pop-canon}. Only the synthetic discs that live beyond \SI{2}{\mega\year} contribute to this histogram. The few remaining discs have low masses, as the discs are close to being dissipated. Using larger values of $\alpha$, for instance between roughly \num{e-3} and \num{e-2} as was proposed by \citet{2017ApJMulders}, would have resulted in even shorter disc lifetimes. This means that there would have been no discs that would live long enough to produce stellar accretion at \SI{2}{\mega\year}. Conversely, selecting a distribution of $\alpha$ with even lower values would allow disc lifetimes to be matched by observations, but this would result in even lower stellar accretion rates, which would be in tension with the results of \citet{2018ApJDullemond}, who concluded that $\alpha\geq\num{1e-4}$ from the sizes of disc substructures.

The discrepancy between our modelled lifetimes and observations arises from the strong mass-loss rates predicted for internal and external photoevaporation, as we discuss in Sect.~\ref{sec:discussion}. As disc lifetime depends on stellar mass (especially for external photoevaporation; Sect.~\ref{sec:res-photo}), this analysis is affected by the assumed stellar mass distribution. Had we selected larger stellar masses, the lifetimes would better match observations. However, selecting stellar masses of around \SI{1}{\msol}, which would lead to a fairly good match including both photoevaporation prescriptions, is not representative of the star-forming regions that we are  comparing to (Sect.~\ref{sec:obs-mstar}).

\subsection{Effect of photoevaporation on disc populations}

The mismatch in disc lifetimes and disc masses that we obtained is in contrast with other studies, such as that of \citet{2017ApJMulders}, who did not include photoevaporation, \citet{2020MNRASKunitomo}, who included only internal photoevaporation, and \citet{Weder2023}, who used an EUV-only internal photoevaporation prescription with much lower mass-loss rates. Further, as already discussed, photoevaporation leads to considerable shortening of disc lifetimes (Sect.~\ref{sec:res-photo}). Therefore, here we investigate how the photoevaporation prescriptions affect the evolution of disc populations and determine whether they are responsible for the mismatch. To this end, we created two additional populations, each time neglecting one of the photoevaporation process by taken the corresponding controlling parameter to the minimum value. For the population with internal photoevaporation only, the ambient flux has been set to $F=1 G_0$, which corresponds to $\dot{M}=\SI{1e-15}{\msol\per\year}$ (Sect.~\ref{sec:method-ext-phew}). The results are shown as `Int. only' in Figs.~\ref{fig:dlt-pop-canon} and~\ref{fig:mar-pop-canon}. For the population with external photoevaporation only, we set $L_\mathrm{X}=\SI{e28}{\erg\per\second}$, the results of which are shown as `Ext. only' in the same figures.

In each population, the disc lifetimes strongly increase compared to the canonical case. However, the distributions are different: internal photoevaporation leads to a relatively narrow distribution around \num{2} to \SI{3}{\mega\year}, while with external photoevaporation disc lifetimes are more spread out, including very short-lived discs. This leads to more discs being shown in the accretion versus mass diagram.

The population with only external photoevaporation has low disc masses combined with a narrow range of stellar accretion between \num{e-10} and \SI{e-9}{\msol\per\year}. This is because the mass loss occurs in the outer disc, which reduces its size, limiting the area from which the dust emits (as dust is also lost where there is no longer gas present). At the same time, the mass-accretion rate is weakly affected by the mass loss in the outer disc; again because external photoevaporation affects only the outer disc.

Conversely, using internal photoevaporation leads to a larger spread in stellar accretion rate. As internal photoevaporation removes material relatively close to the star, it competes with stellar accretion to some extent \citep{2020MNRASSomigliana}. This, coupled with the spread of X-ray luminosities, leads to a spread in accretion rate \citep{2011MNRASOwen}. Thus, internal photoevaporation is needed to reproduce the observed spread in stellar accretion rate \citep{2016AAManaraB,2022AATesti}. We further note that neither population is able to reproduce the discs with an accretion rate larger than \SI{e-8}{\msol\per\year}.

In addition, internal photoevaporation leaves discs with an inner cavity that have low accretion rates but where the outer disc is out of reach of internal photoevaporation and therefore takes a long time to dissipate; these are what \citet{2011MNRASOwen} refer to as relic discs. It is possible to avoid this situation to a large extent by having more compact discs initially, which leave nearly all of their mass within reach of internal photoevaporation.

We conclude that previous studies managed to reproduce disc lifetimes because they used only one photoevaporation mechanism as the main loss mechanism. However, when both are accounted for, the combined mass-loss rate is so large that discs are very short lived. We discuss the implications of this and possible remedies in Sect.~\ref{sec:discussion}.

\subsection{Towards a best match}

\begin{figure}
	\centering
	\includegraphics{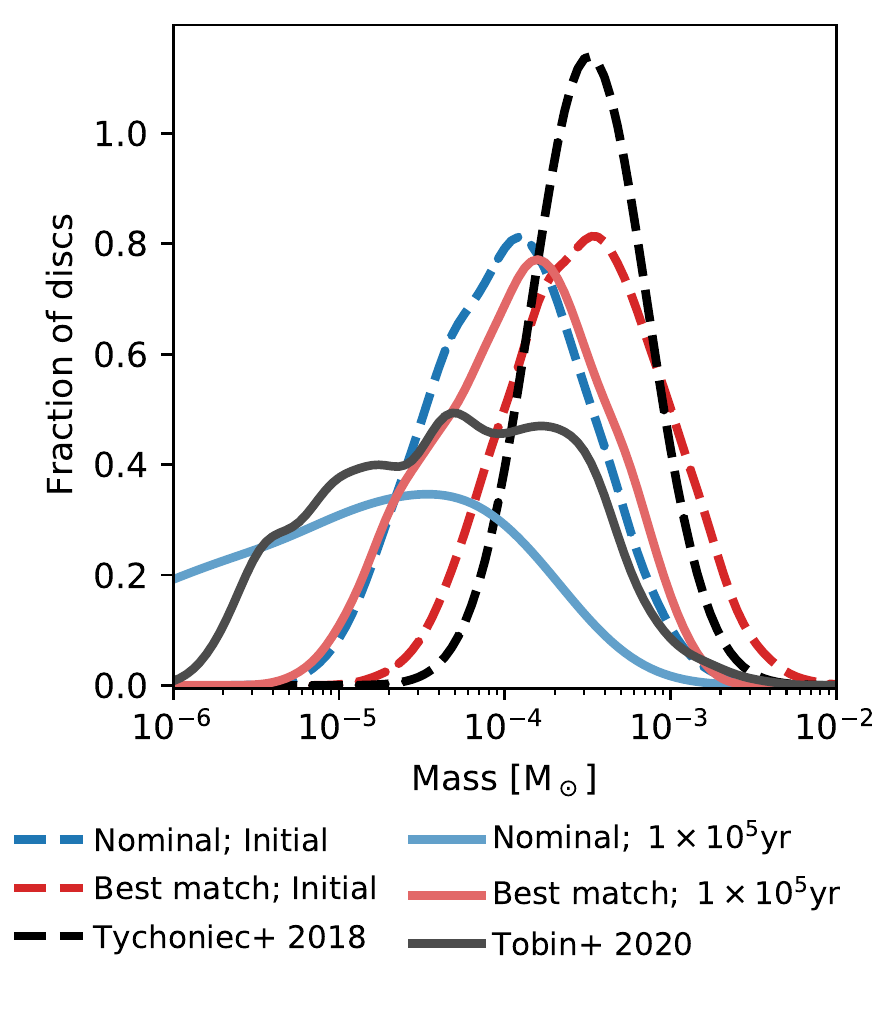}
	\caption{Kernel density estimate for two populations from Figs.~\ref{fig:dlt-pop-canon} and~\ref{fig:mar-pop-canon}, both for their initial conditions (dashed lines) and retrieved disc masses at \SI{100}{kyr} (solid lines). The initial mass distribution of the best-match population with $M_\star^\mathrm{ref}=\SI{0.33}{\msol}$ (in red) is consistent with the fit by \citet{2021AAEmsenhuberB} to the data of \citet{2018ApJSTychoniec}, which is shown as the dashed black line, while the retrieved masses at \SI{100}{\kilo\year} are compatible with the non-multiple discs of \citet{2020ApJTobinA}.}
	\label{fig:mass-canon}
\end{figure}

Despite all the differences between models and observations highlighted so far, we now try to find a set of initial conditions that is able to better match disc evolution characteristics. To this end, we investigate how the initial conditions can be modified from our canonical values provided in Table~\ref{tab:canon-vars}.

The initial disc-to-star mass ratios were taken from \citet{2018ApJSTychoniec} and \citet{2020ApJTobinA} assuming they were measured on stars of $M_\star^\mathrm{ref}=\SI{1}{\msol}$. However, this assumption is inconsistent with the stellar mass distribution we selected, which has a median value of $M_\star=\SI{0.35}{\msol}$ (Sect.~\ref{sec:obs-mstar}). Were we to select a lower reference stellar mass, such as $M_\star^\mathrm{ref}=\SI{0.33}{\msol}$, this would increase the disc masses. At the same time, selecting a reference stellar mass that is similar to the median value from our initial conditions results in an agreement between the disc masses in observations and our initial conditions, as shown with the dashed black and red lines in Fig.~\ref{fig:mass-canon}, respectively. In addition, to account for observational biases, we also want to check the observed dust masses after a short evolution time of \SI{1e5}{\year}, which we provide with the solid lines in Fig.~\ref{fig:mass-canon}. The results show that the discs in the nominal populations have low masses compared to the non-multiple discs measured by \citet{2020ApJTobinA}, while a population with more massive initial discs has slightly larger masses. Thus, the larger initial disc masses lead to a reasonable match with observations as a whole.

We point out that the disc-to-star mass ratio in the population with $M_\star^\mathrm{ref}=\SI{0.33}{\msol}$ is about twice ($2.1$  times) that of \citet{2021AAEmsenhuberB} and \citet{2021AABurn}, rather than a factor three as one might assume from the change of $M_\star^\mathrm{ref}$ from \SI{1}{\msol} to \SI{0.33}{\msol}. This is due to an inconsistency in the selection of the disc masses in the previous work: there, the gas masses were taken as a Monte Carlo variable that were converted from dust observations using a dust-to-gas ratio of \SI{1}{\percent}. However, the initial mass of solids in the model was recomputed from the gas mass using the same distribution as in this work (Sect.~\ref{sec:obs-fdg}), which has a median value of \SI{1.42}{\percent} ($f_\mathrm{D/G,\odot}=\SI{1.49}{\percent}$ with $\mathrm{[Fe/H]}=-0.02$). In contrast, we use the solid disc mass as a Monte Carlo variable here.

Another possibility is that early protoplanetary discs are not as extended as what is suggested by the findings of \citet{2020ApJTobinA}. More compact discs are less susceptible to external photoevaporation, as there is less surface exposed to ambient radiation and they are more tightly bound to the star. At the same time, a more compact disc means that more material is concentrated in the region where internal photoevaporation is most efficient, which allows the stellar accretion rate to remain larger. Magnetohydrodynamics models of protoplanetary disc formation by \citet{2016ApJHennebelle}, \citet{2021ApJLebreuilly}, and \citet{2021ApJLee} could favour such a possibility.

Finally, nearby star-forming regions have low masses, which results in a low ambient UV field strength $F$; for instance, the value in Lupus is $F \approx 4G_0$ \citep{2016ApJCleevesC}. Our nominal distribution of ambient UV fields overestimates mass losses due to external photoevaporation. Therefore, we set $F=1G_0$, which results in negligible mass losses due to external photoevaporation ($\dot{M}=\SI{1e-15}{\msol\per\year}$) and only internal photoevaporation remains. Using internal photoevaporation confers the further advantage that a wider range of stellar accretion rates is obtained.

Below, we investigate whether more massive and compact discs are able to improve the match with observations. The effects of the parameters mentioned above are discussed in Appendix~\ref{sec:popparams}. From these, we find that the following modifications to the initial conditions given in Table~\ref{tab:canon-vars} are best able to reproduce disc lifetimes and the accretion rate--mass relationship:
\begin{itemize}
    \item A decrease in the reference stellar mass to $M_\star^\mathrm{ref}=\SI{0.33}{\msol}$, which corresponds to an increase in the disc mass by a factor of three compared to the nominal population;
    \item $r_1=2/3\times\SI{70}{au}\left(M_\mathrm{D}/\SI{100}{\mearth}\right)^{0.25}$, a factor $2/3$ compared to Eq.~(\ref{eq:size}); and
    \item only using internal photoevaporation ($F=1G_0$).
\end{itemize}
The population using these distributions is shown as `Best match' in Figs.~\ref{fig:dlt-pop-canon} and~\ref{fig:mar-pop-canon}.

While the overall stellar mass distribution we chose is representative of the observed stars, our visual optimisation approach to reproducing the observed disc masses and accretion rates is independent of the exact stellar mass dependency of the observables. We will improve on this with a Bayesian framework in \paperthree{} to optimise the initial parameters when reproducing a set of observations in four-dimensional space made up of stellar mass, disc mass, disc radius, and accretion rate.

Compared to the population with only internal photoevaporation (the other population that is closest in terms of initial conditions), we can see several differences. First, the larger disc masses result in an increased median lifetime. About \SI{2.2}{\percent} of the discs have a lifetime of greater than \SI{10}{\mega\year}, representing a 100-fold increase. Then, the combination of larger initial disc mass and smaller extent results in a certain number of cases with a stellar accretion rate of higher than \SI{e-8}{\msol\per\year}, which was not previously seen. The smaller extent of the disc also causes less discs to be relics (towards the bottom of Fig.~\ref{fig:mar-pop-canon}). Here, the highest concentration of discs is found near the best-fit value of \citet[the pink dashed line in Fig.~\ref{fig:mar-pop-canon}]{2016AAManaraB} with a similar number on either side. We are still failing to reproduce the discs with the large accretion rates. However, part of these discs with large accretion rates could be due to binaries \citep{2022MNRASZagariaA}, which we do not model. To corroborate this, the largest stellar accretion rates are biased to larger-mass stars (the mean stellar mass for systems with a stellar accretion rate higher than \SI{3e-9}{\msol\per\year} is \SI{0.74}{\msol} versus \SI{0.43}{\msol} for the general population), which at the same time are more likely to be in binary systems \citep[e.g.][]{2013ARAADucheneKraus}. The mismatch should therefore not be the source of significant concern. We find that this combination of parameters is able to reproduce the disc mass--accretion rate relationship and provide a reasonable match to most observations.

\section{Discussion}
\label{sec:discussion}

\begin{figure}
	\centering
	\includegraphics{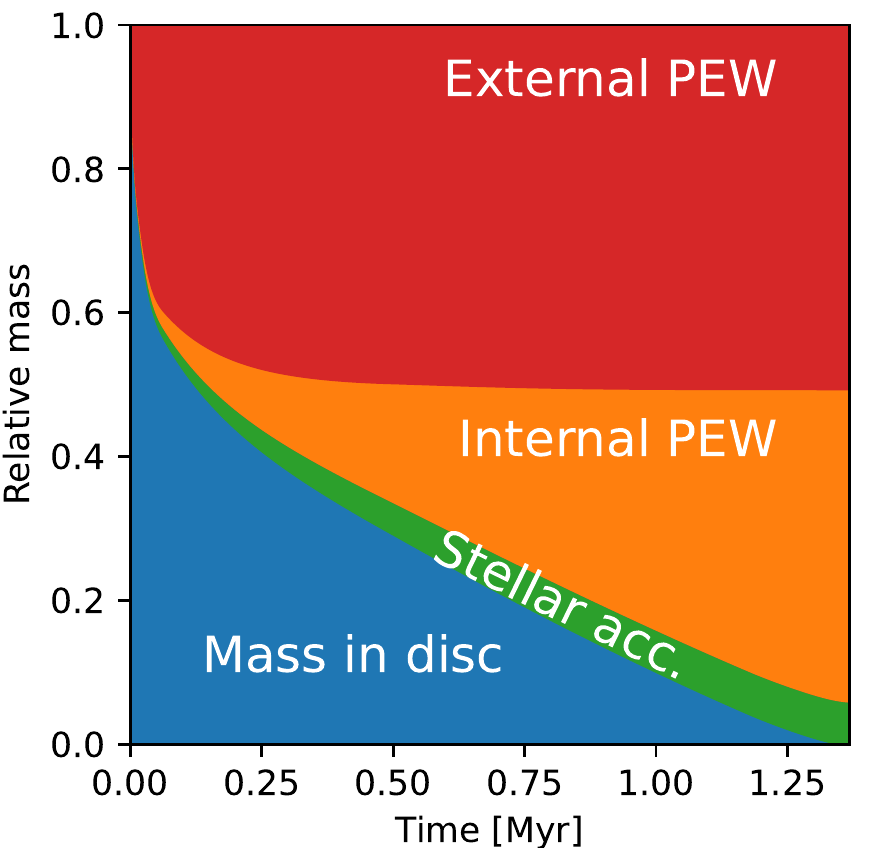}
	\caption{Evolution of the relative gas mass: in the disc (blue), accreted onto the star (green), and lost by internal (orange) or external (red) photoevaporation until disc dispersal at \SI{1.38}{\mega\year}. This represents a typical disc, with $M_\star=\SI{0.5}{\msol}$, $M_\mathrm{G}/M_\star=\num{0.1}$, $\beta=0.9$, $R_\mathrm{in}=\SI{0.1}{\au}$, $r_1=\SI{87.8}{\au}$, $\alpha=\num{1e-3}$, $L_\mathrm{X}=\SI{7.02e29}{\erg\per\second}$, and $F=\SI{10}{G_0}$. The parameters for the solid disc are irrelevant, as this shows only the gas component, except that we used $f_\mathrm{D/G}=\num{0.0149}$ to compute the solid mass needed to set the characteristic radius $r_1$.}
	\label{fig:typical}
\end{figure}

It is difficult to reconcile the results of our synthetic disc populations with observations. We find that it is particularly hard to obtain discs with characteristic lifetimes of \num{2} to \SI{3}{\mega\year} according to \citet{2009AIPCMamajek} or \citet{2010AAFedele}, and even less lifetimes of \num{5} to \SI{10}{\mega\year} following \citet{2022ApJPfalzner}. Here, we assumed that the initial mass is constrained by the dust-mass measurements of \citet[$M_\mathrm{D}/M_\star\sim\num{e-3}$]{2018ApJSTychoniec} , and dust-to-gas ratios similar to the solar initial abundance, namely $f_\mathrm{D/G}=\SI{1.49}{\percent}$ \citep{2003ApJLodders}, combined with the predictions of internal and external photoevaporation models. To illustrate this issue, in Fig.~\ref{fig:typical}  we provide the evolution of the gas mass still present in the disc and removed by the processes modelled here. Here we choose typical values for the initial conditions, with a star of mass $M_\star=\SI{0.5}{\msol}$, a gas disc with a mass of $M_\mathrm{G}/M_\star=\num{0.1}$, a power-law index of $\beta=0.9$, an inner radius of $r_\mathrm{in}=\SI{0.1}{\au}$, and a characteristic radius of $r_1=\SI{87.8}{au}$, which was computed from Eq.~(\ref{eq:size}) assuming a dust-to-gas ratio of $f_\mathrm{D/G}=\num{0.0149}$. The ambient UV field strength $F=\SI{10}{G_0}$ was set at the lower boundary of the computed grid while the X-ray luminosity $L_\mathrm{X}=\SI{7.02e29}{\erg\per\second}$ follows the best fit of the XEST survey \citep[Table~\ref{tab:lx-fits}]{2007AAGudel} for the given stellar mass.

Figure~\ref{fig:typical} shows that photoevaporation (both internal and external) is responsible for the loss of nearly all the gas; only \SI{6}{\percent} of the gas is accreted onto the star. The final disc lifetime of \SI{1.38}{\mega\year} is short compared to the characteristic lifetimes from observations. We therefore have a problem with the mass budget, which could be resolved by (i) larger initial disc masses, (ii) mass replenishment after disc formation, or (iii) lower mass-loss rates by photoevaporation.

Stellar accretion already plays a minor role in the mass budget and cannot be reduced further, in order to remain consistent with observed stellar accretion rates. However, radial mass transport could be the result of magnetically driven disc winds \citep{2009ApJSuzukiInutsuka,2010ApJSuzuki} rather than pure viscous dissipation as we assumed here. This would add another mass-loss channel, which would further exacerbate the problem (though it could shield stellar radiation, as discussed below).

We experiment with larger initial disc masses in this work. However, even with gas masses on the order of \SI{10}{\percent} of the stellar mass, disc lifetimes are not sufficiently long. Therefore, increasing the masses even more would be required, but this increase would bring another series of problems. For one, such large discs are likely gravitationally unstable and produce spiral density waves. Second, such large discs would lead to strong gas-driven migration, which hinders planet formation \citep{2022MNRASNayakshin}. Also, discs around \SI{10}{\mega\year}-old stars HD163296 and TW Hya have at least \SI{10}{\percent} of the stellar mass \citep{2019ApJPowell} and it is unclear what their initial mass  would have been for them to remain so large at their age. We therefore do not find that massive initial discs would be able to solve the conundrum.

Discs do not form instantly. Rather, they grow from gas falling from the envelope. This allows for disc masses to increase during the early stages of disc evolution, which is at odds with the models described here. Modelling the infall stage would allow us to have longer-lived discs without them being very massive early on. Accretion can persist for several million  years \citep{2008AJThroopBally}, providing replenishment even at late times. The complex morphology of the gas disc around RU Lup (a \SI{\sim0.5}{\mega\year} old star) could be an outcome of this process \citep{2020ApJHuang}. The long-lived discs could also be second-generation discs, formed following accretion from the molecular cloud \citep{2020AAKuffmeierA} or the disruption of a planet \citep{2020MNRASNayakshinB}. While none of these individual processes are sufficient to explain the presence of massive discs, they should still be explored if they can explain certain characteristics of the overall disc population.

Finally, the photoevaporation rates used here could be overestimated. Lower mass-loss rates would increase the disc lifetimes and masses at later times. An argument in favour of this hypothesis is that young discs can be shielded from the radiation of both their host and nearby stars. The launching of magnetically driven disc winds occurs inside the location where internal photoevaporation is effective and would thus prevent EUV and X-ray photoevaporation during the early stages of disc evolution \citep{2022PPVIIPascucci}. Similarly, early discs are likely embedded, preventing the radiation of nearby stars from reaching the disc. Both effects would reduce the photoevaporation rates during the early stage of disc evolution compared to what we model here.

\section{Summary and conclusion}

In this work, we investigate whether protoplanetary disc observations can be reconciled with theoretical predictions of processes such as viscous accretion and photoevaporation (both internal and external). We first compute two sets of simulations that we use to fit neural networks (Fig.~\ref{fig:corr-dlt-nir}). With these neural networks, we can perform parameter studies and compute the outcomes of synthetic disc populations with limited computational resources.

We first compare how internal and external photoevaporation affect disc lifetime as a function of stellar mass. We find that because of a direct link between stellar mass and X-ray luminosity \citep[e.g.][]{2005ApJSPreibischA,2007AAGudel}, which means mass-loss rate due to internal photoevaporation, discs around more massive stars are not significantly longer-lived than  those around low-mass stars (Fig.~\ref{fig:int-dlt-nir}). Conversely, external photoevaporation leads to a strong positive correlation between disc lifetime and stellar mass (Fig.~\ref{fig:ext-dlt-nir}), because gas is more bound for higher-mass stars. This positive correlation is at odds with observations that find that disc lifetimes are either independent of or anticorrelated with stellar mass \citep{2006ApJCarpenter,2009ApJKennedyKenyon,2012AABayo,2015AARibas,2022ApJPfalzner} and should be investigated in the future.

Turning to protoplanetary disc populations, we find that accounting for both internal and external photoevaporation according to theoretical predictions leads to disc lifetimes (Fig.~\ref{fig:dlt-pop-canon}) that are much too short. Discs whose initial mass is \SI{10}{\percent} of the stellar mass are dispersed in roughly \SI{1}{\mega\year} under the combined effects of internal and external photoevaporation (Fig. \ref{fig:typical}).

Despite the dissimilarities, a reasonably good match to the disc properties of the Lupus and Chaemeleon~I low-mass star-forming regions is obtained starting with more massive discs of smaller sizes, and with only internal photoevaporation. This is valid for clusters with low ambient field strengths, such as Lupus (\SI{4}{G_0}; \citealp{2016ApJCleevesC}), where losses due to external photoevaporation are low. The larger masses and smaller sizes are needed to improve the match in stellar accretion rates and observed masses. The corresponding initial conditions and model parameters can be used to study planetary formation in similar environments. A more robust comparison with observations is performed in \paperthree.

However, initial disc masses cannot be arbitrarily increased or discs would become gravitationally unstable. Instead, we suggest that future studies should include the modelling of the initial stages of disc formation, including the presence of an envelope. This envelope would allow discs to be replenished after their initial formation and provide shielding from UV radiation from nearby stars. Also, magnetically driven disc winds would shield UV and X-ray radiation from the central star. This would provide a reduction of losses by both internal and external photoevaporation. Both effects allow for longer lifetimes and larger masses at later times without the need for extremely large masses at earlier times.

We decided here to use observed dust masses as the main comparison point. However, this is not the only possible avenue. For instance, disc radii could be less susceptible to the degeneracy caused by regions that are optically thick \citep{2022PPVIIPascucci}. One likely difficulty would be in accounting for the large disc radii and sustained stellar accretion rate. Already with the comparatively small discs that we find to best match the disc mass--stellar accretion rate relationship, we are not able to reproduce the largest observed accretion rates. Having larger discs would lead to a reduction of the stellar accretion rates for a given disc mass (Fig.~\ref{fig:mar-params}); which would lead to a mismatch with observations of stellar accretion rates.

In this work, we assume that the gas discs evolve viscously. However, simulations of disc evolution that account for non-ideal magnetohydrodynamical (MHD) effects find that the magnetorotational instability (MRI), which is the likely mechanism generating the turbulence, is largely suppressed \citep[e.g.][]{2013ApJBaiStoneB,2014AALesur}. Instead, it has also been proposed that the evolution is driven by magnetically driven disc winds \citep{2009ApJSuzukiInutsuka}. Magnetically driven disc wind prescriptions, such as those of \citet{2016AASuzuki} or \citet{2022MNRASTaboneA}, include several model possibilities, which can be narrowed down by performing a similar comparison to that presented here \citep{Weder2023}. Once such a model is properly coupled with internal photoevaporation to account for shielding, a similar study to that presented here is possible.

\begin{acknowledgements}
The authors thank Christian Rab, Ilaria Pascucci, Susanne Pfalzner, and Aashish Gupta for fruitful discussions.
We also thank the anonymous reviewer, whose comments and suggestions greatly helped improve the manuscript's quality.
This work was funded by the Deutsche Forschungsgemeinschaft (DFG, German Research Foundation) - 362051796.
This research was supported by the Excellence Cluster ORIGINS which is funded by the Deutsche Forschungsgemeinschaft (DFG, German Research Foundation) under Germany's Excelence Strategy – EXC-2094-390783311.
The plots shown in this work were generated using \textit{matplotlib} \citep{2007CSEHunter}.
\end{acknowledgements}

\bibliographystyle{aa}
\bibliography{manu,add}

\begin{thebibliography}{146}
\expandafter\ifx\csname natexlab\endcsname\relax\def\natexlab#1{#1}\fi

\bibitem[{{Adams} {et~al.}(2006){Adams}, {Proszkow}, {Fatuzzo}, \&
  {Myers}}]{2006ApJAdams}
{Adams}, F.~C., {Proszkow}, E.~M., {Fatuzzo}, M., \& {Myers}, P.~C. 2006, \apj,
  641, 504

\bibitem[{{Alcal{\'a}} {et~al.}(2017){Alcal{\'a}}, {Manara}, {Natta}, {Frasca},
  {Testi}, {Nisini}, {Stelzer}, {Williams}, {Antoniucci}, {Biazzo}, {Covino},
  {Esposito}, {Getman}, \& {Rigliaco}}]{2017AAAlcala}
{Alcal{\'a}}, J.~M., {Manara}, C.~F., {Natta}, A., {et~al.} 2017, \aap, 600,
  A20

\bibitem[{{Alcal{\'a}} {et~al.}(2014){Alcal{\'a}}, {Natta}, {Manara}, {Spezzi},
  {Stelzer}, {Frasca}, {Biazzo}, {Covino}, {Randich}, {Rigliaco}, {Testi},
  {Comer{\'o}n}, {Cupani}, \& {D'Elia}}]{2014AAAlcala}
{Alcal{\'a}}, J.~M., {Natta}, A., {Manara}, C.~F., {et~al.} 2014, \aap, 561, A2

\bibitem[{{Alexander} {et~al.}(2004){Alexander}, {Clarke}, \&
  {Pringle}}]{2004MNRASAlexander}
{Alexander}, R.~D., {Clarke}, C.~J., \& {Pringle}, J.~E. 2004, \mnras, 354, 71

\bibitem[{{Alibert} {et~al.}(2004){Alibert}, {Mordasini}, \&
  {Benz}}]{2004A&AAlibert}
{Alibert}, Y., {Mordasini}, C., \& {Benz}, W. 2004, \aap, 417, L25

\bibitem[{{Alibert} {et~al.}(2005){Alibert}, {Mordasini}, {Benz}, \&
  {Winisdoerffer}}]{2005A&AAlibert}
{Alibert}, Y., {Mordasini}, C., {Benz}, W., \& {Winisdoerffer}, C. 2005, \aap,
  434, 343

\bibitem[{{Andrews} {et~al.}(2018{\natexlab{a}}){Andrews}, {Huang},
  {P{\'e}rez}, {Isella}, {Dullemond}, {Kurtovic}, {Guzm{\'a}n}, {Carpenter},
  {Wilner}, {Zhang}, {Zhu}, {Birnstiel}, {Bai}, {Benisty}, {Hughes},
  {{\"O}berg}, \& {Ricci}}]{2018ApJAndrewsB}
{Andrews}, S.~M., {Huang}, J., {P{\'e}rez}, L.~M., {et~al.} 2018{\natexlab{a}},
  \apjl, 869, L41

\bibitem[{{Andrews} {et~al.}(2018{\natexlab{b}}){Andrews}, {Terrell},
  {Tripathi}, {Ansdell}, {Williams}, \& {Wilner}}]{2018ApJAndrewsA}
{Andrews}, S.~M., {Terrell}, M., {Tripathi}, A., {et~al.} 2018{\natexlab{b}},
  \apj, 865, 157

\bibitem[{{Andrews} {et~al.}(2009){Andrews}, {Wilner}, {Hughes}, {Qi}, \&
  {Dullemond}}]{2009ApJAndrews}
{Andrews}, S.~M., {Wilner}, D.~J., {Hughes}, A.~M., {Qi}, C., \& {Dullemond},
  C.~P. 2009, \apj, 700, 1502

\bibitem[{{Andrews} {et~al.}(2010){Andrews}, {Wilner}, {Hughes}, {Qi}, \&
  {Dullemond}}]{2010ApJAndrews}
{Andrews}, S.~M., {Wilner}, D.~J., {Hughes}, A.~M., {Qi}, C., \& {Dullemond},
  C.~P. 2010, \apj, 723, 1241

\bibitem[{{Ansdell} {et~al.}(2017){Ansdell}, {Williams}, {Manara}, {Miotello},
  {Facchini}, {van der Marel}, {Testi}, \& {van Dishoeck}}]{2017AJAnsdell}
{Ansdell}, M., {Williams}, J.~P., {Manara}, C.~F., {et~al.} 2017, \aj, 153, 240

\bibitem[{{Ansdell} {et~al.}(2016){Ansdell}, {Williams}, {van der Marel},
  {Carpenter}, {Guidi}, {Hogerheijde}, {Mathews}, {Manara}, {Miotello},
  {Natta}, {Oliveira}, {Tazzari}, {Testi}, {van Dishoeck}, \& {van
  Terwisga}}]{2016ApJAnsdell}
{Ansdell}, M., {Williams}, J.~P., {van der Marel}, N., {et~al.} 2016, \apj,
  828, 46

\bibitem[{{Bai} \& {Stone}(2013)}]{2013ApJBaiStoneB}
{Bai}, X.-N. \& {Stone}, J.~M. 2013, \apj, 769, 76

\bibitem[{{Baraffe} {et~al.}(1998){Baraffe}, {Chabrier}, {Allard}, \&
  {Hauschildt}}]{1998AABaraffe}
{Baraffe}, I., {Chabrier}, G., {Allard}, F., \& {Hauschildt}, P.~H. 1998, \aap,
  337, 403

\bibitem[{{Baraffe} {et~al.}(2015){Baraffe}, {Homeier}, {Allard}, \&
  {Chabrier}}]{2015AABaraffe}
{Baraffe}, I., {Homeier}, D., {Allard}, F., \& {Chabrier}, G. 2015, \aap, 577,
  A42

\bibitem[{{Bayo} {et~al.}(2012){Bayo}, {Barrado}, {Hu{\'e}lamo},
  {Morales-Calder{\'o}n}, {Melo}, {Stauffer}, \& {Stelzer}}]{2012AABayo}
{Bayo}, A., {Barrado}, D., {Hu{\'e}lamo}, N., {et~al.} 2012, \aap, 547, A80

\bibitem[{{Beckwith} \& {Sargent}(1996)}]{1996NatureBeckwithSargent}
{Beckwith}, S. V.~W. \& {Sargent}, A.~I. 1996, \nat, 383, 139

\bibitem[{{Birnstiel} {et~al.}(2010){Birnstiel}, {Dullemond}, \&
  {Brauer}}]{2010AABirnstiel}
{Birnstiel}, T., {Dullemond}, C.~P., \& {Brauer}, F. 2010, \aap, 513, A79

\bibitem[{{Birnstiel} {et~al.}(2018){Birnstiel}, {Dullemond}, {Zhu}, {Andrews},
  {Bai}, {Wilner}, {Carpenter}, {Huang}, {Isella}, {Benisty}, {P{\'e}rez}, \&
  {Zhang}}]{2018ApJBirnstiel}
{Birnstiel}, T., {Dullemond}, C.~P., {Zhu}, Z., {et~al.} 2018, \apjl, 869, L45

\bibitem[{{Birnstiel} {et~al.}(2012){Birnstiel}, {Klahr}, \&
  {Ercolano}}]{2012A&ABirnstiel}
{Birnstiel}, T., {Klahr}, H., \& {Ercolano}, B. 2012, \aap, 539, A148

\bibitem[{Burn {et~al.}(in prep.)Burn, Emsenhuber, Melon~Fuksman, Henning,
  Ercolano, \& Klahr}]{Disk3}
Burn, R., Emsenhuber, A., Melon~Fuksman, J.~D., {et~al.} in prep., \aap

\bibitem[{{Burn} {et~al.}(2022){Burn}, {Emsenhuber}, {Weder}, {V{\"o}lkel},
  {Klahr}, {Birnstiel}, {Ercolano}, \& {Mordasini}}]{2022AABurnA}
{Burn}, R., {Emsenhuber}, A., {Weder}, J., {et~al.} 2022, \aap, 666, A73

\bibitem[{{Burn} {et~al.}(2021){Burn}, {Schlecker}, {Mordasini}, {Emsenhuber},
  {Alibert}, {Henning}, {Klahr}, \& {Benz}}]{2021AABurn}
{Burn}, R., {Schlecker}, M., {Mordasini}, C., {et~al.} 2021, \aap, 656, A72

\bibitem[{Byrd {et~al.}(1995)Byrd, Lu, Nocedal, \& Zhu}]{1995SIAMJSCByrd}
Byrd, R.~H., Lu, P., Nocedal, J., \& Zhu, C. 1995, SIAM Journal on Scientific
  Computing, 16, 1190

\bibitem[{{Cambioni} {et~al.}(2019{\natexlab{a}}){Cambioni}, {Asphaug},
  {Emsenhuber}, {Gabriel}, {Furfaro}, \& {Schwartz}}]{2019ApJCambioni}
{Cambioni}, S., {Asphaug}, E., {Emsenhuber}, A., {et~al.} 2019{\natexlab{a}},
  \apj, 875, 40

\bibitem[{{Cambioni} {et~al.}(2019{\natexlab{b}}){Cambioni}, {Delbo}, {Ryan},
  {Furfaro}, \& {Asphaug}}]{2019IcarusCambioni}
{Cambioni}, S., {Delbo}, M., {Ryan}, A.~J., {Furfaro}, R., \& {Asphaug}, E.
  2019{\natexlab{b}}, \icarus, 325, 16

\bibitem[{{Cambioni} {et~al.}(2021){Cambioni}, {Jacobson}, {Emsenhuber},
  {Asphaug}, {Rubie}, {Gabriel}, {Schwartz}, \& {Furfaro}}]{2021PSJCambioni}
{Cambioni}, S., {Jacobson}, S.~A., {Emsenhuber}, A., {et~al.} 2021, \psj, 2, 93

\bibitem[{{Carpenter} {et~al.}(2006){Carpenter}, {Mamajek}, {Hillenbrand}, \&
  {Meyer}}]{2006ApJCarpenter}
{Carpenter}, J.~M., {Mamajek}, E.~E., {Hillenbrand}, L.~A., \& {Meyer}, M.~R.
  2006, \apjl, 651, L49

\bibitem[{{Chabrier}(2003)}]{2003PASPChabrier}
{Chabrier}, G. 2003, \pasp, 115, 763

\bibitem[{{Clarke} {et~al.}(2001){Clarke}, {Gendrin}, \&
  {Sotomayor}}]{2001MNRASClarke}
{Clarke}, C.~J., {Gendrin}, A., \& {Sotomayor}, M. 2001, \mnras, 328, 485

\bibitem[{{Cleeves} {et~al.}(2016){Cleeves}, {{\"O}berg}, {Wilner}, {Huang},
  {Loomis}, {Andrews}, \& {Czekala}}]{2016ApJCleevesC}
{Cleeves}, L.~I., {{\"O}berg}, K.~I., {Wilner}, D.~J., {et~al.} 2016, \apj,
  832, 110

\bibitem[{{Da Rio} {et~al.}(2012){Da Rio}, {Robberto}, {Hillenbrand},
  {Henning}, \& {Stassun}}]{2012ApJDaRio}
{Da Rio}, N., {Robberto}, M., {Hillenbrand}, L.~A., {Henning}, T., \&
  {Stassun}, K.~G. 2012, \apj, 748, 14

\bibitem[{{Dittrich} {et~al.}(2013){Dittrich}, {Klahr}, \&
  {Johansen}}]{2013ApJDittrich}
{Dittrich}, K., {Klahr}, H., \& {Johansen}, A. 2013, \apj, 763, 117

\bibitem[{{Dr{\k{a}}{\.z}kowska} \& {Alibert}(2017)}]{2017AADrazkowskaAlibert}
{Dr{\k{a}}{\.z}kowska}, J. \& {Alibert}, Y. 2017, \aap, 608, A92

\bibitem[{{Duch{\^e}ne} \& {Kraus}(2013)}]{2013ARAADucheneKraus}
{Duch{\^e}ne}, G. \& {Kraus}, A. 2013, \araa, 51, 269

\bibitem[{{Dullemond} {et~al.}(2018){Dullemond}, {Birnstiel}, {Huang},
  {Kurtovic}, {Andrews}, {Guzm{\'a}n}, {P{\'e}rez}, {Isella}, {Zhu}, {Benisty},
  {Wilner}, {Bai}, {Carpenter}, {Zhang}, \& {Ricci}}]{2018ApJDullemond}
{Dullemond}, C.~P., {Birnstiel}, T., {Huang}, J., {et~al.} 2018, \apjl, 869,
  L46

\bibitem[{{Emsenhuber} {et~al.}(2020){Emsenhuber}, {Cambioni}, {Asphaug},
  {Gabriel}, {Schwartz}, \& {Furfaro}}]{2020ApJEmsenhuberA}
{Emsenhuber}, A., {Cambioni}, S., {Asphaug}, E., {et~al.} 2020, \apj, 891, 6

\bibitem[{{Emsenhuber} {et~al.}(2021{\natexlab{a}}){Emsenhuber}, {Mordasini},
  {Burn}, {Alibert}, {Benz}, \& {Asphaug}}]{2021AAEmsenhuberA}
{Emsenhuber}, A., {Mordasini}, C., {Burn}, R., {et~al.} 2021{\natexlab{a}},
  \aap, 656, A69

\bibitem[{{Emsenhuber} {et~al.}(2021{\natexlab{b}}){Emsenhuber}, {Mordasini},
  {Burn}, {Alibert}, {Benz}, \& {Asphaug}}]{2021AAEmsenhuberB}
{Emsenhuber}, A., {Mordasini}, C., {Burn}, R., {et~al.} 2021{\natexlab{b}},
  \aap, 656, A70

\bibitem[{{Ercolano} {et~al.}(2009){Ercolano}, {Clarke}, \&
  {Drake}}]{2009ApJErcolano}
{Ercolano}, B., {Clarke}, C.~J., \& {Drake}, J.~J. 2009, \apj, 699, 1639

\bibitem[{{Ercolano} {et~al.}(2011){Ercolano}, {Clarke}, \&
  {Hall}}]{2011MNRASErcolano}
{Ercolano}, B., {Clarke}, C.~J., \& {Hall}, A.~C. 2011, \mnras, 410, 671

\bibitem[{{Ercolano} {et~al.}(2008){Ercolano}, {Drake}, {Raymond}, \&
  {Clarke}}]{2008ApJErcolano}
{Ercolano}, B., {Drake}, J.~J., {Raymond}, J.~C., \& {Clarke}, C.~C. 2008,
  \apj, 688, 398

\bibitem[{{Ercolano} {et~al.}(2015){Ercolano}, {Koepferl}, {Owen}, \&
  {Robitaille}}]{2015MNRASErcolano}
{Ercolano}, B., {Koepferl}, C., {Owen}, J., \& {Robitaille}, T. 2015, \mnras,
  452, 3689

\bibitem[{{Ercolano} {et~al.}(2021){Ercolano}, {Picogna}, {Monsch}, {Drake}, \&
  {Preibisch}}]{2021MNRASErcolano}
{Ercolano}, B., {Picogna}, G., {Monsch}, K., {Drake}, J.~J., \& {Preibisch}, T.
  2021, \mnras, 508, 1675

\bibitem[{{Facchini} {et~al.}(2016){Facchini}, {Clarke}, \&
  {Bisbas}}]{2016MNRASFacchini}
{Facchini}, S., {Clarke}, C.~J., \& {Bisbas}, T.~G. 2016, \mnras, 457, 3593

\bibitem[{{Favata} \& {Micela}(2003)}]{2003SSRvFavataMicela}
{Favata}, F. \& {Micela}, G. 2003, \ssr, 108, 577

\bibitem[{{Fedele} {et~al.}(2010){Fedele}, {van den Ancker}, {Henning},
  {Jayawardhana}, \& {Oliveira}}]{2010AAFedele}
{Fedele}, D., {van den Ancker}, M.~E., {Henning}, T., {Jayawardhana}, R., \&
  {Oliveira}, J.~M. 2010, \aap, 510, A72

\bibitem[{{Feigelson} \& {Montmerle}(1999)}]{1999ARAAFeigelsonMontmerle}
{Feigelson}, E.~D. \& {Montmerle}, T. 1999, \araa, 37, 363

\bibitem[{{Flaischlen} {et~al.}(2021){Flaischlen}, {Preibisch}, {Manara}, \&
  {Ercolano}}]{2021AAFlaischlen}
{Flaischlen}, S., {Preibisch}, T., {Manara}, C.~F., \& {Ercolano}, B. 2021,
  \aap, 648, A121

\bibitem[{{Fortier} {et~al.}(2013){Fortier}, {Alibert}, {Carron}, {Benz}, \&
  {Dittkrist}}]{2013A&AFortier}
{Fortier}, A., {Alibert}, Y., {Carron}, F., {Benz}, W., \& {Dittkrist}, K.~M.
  2013, \aap, 549, A44

\bibitem[{{Franz} {et~al.}(2020){Franz}, {Picogna}, {Ercolano}, \&
  {Birnstiel}}]{2020AAFranz}
{Franz}, R., {Picogna}, G., {Ercolano}, B., \& {Birnstiel}, T. 2020, \aap, 635,
  A53

\bibitem[{{G{\'a}rate} {et~al.}(2020){G{\'a}rate}, {Birnstiel},
  {Dr{\k{a}}{\.z}kowska}, \& {Stammler}}]{2020AAGarate}
{G{\'a}rate}, M., {Birnstiel}, T., {Dr{\k{a}}{\.z}kowska}, J., \& {Stammler},
  S.~M. 2020, \aap, 635, A149

\bibitem[{{G{\'a}sp{\'a}r} {et~al.}(2016){G{\'a}sp{\'a}r}, {Rieke}, \&
  {Ballering}}]{2016ApJGaspar}
{G{\'a}sp{\'a}r}, A., {Rieke}, G.~H., \& {Ballering}, N. 2016, \apj, 826, 171

\bibitem[{{Getman} {et~al.}(2005){Getman}, {Flaccomio}, {Broos}, {Grosso},
  {Tsujimoto}, {Townsley}, {Garmire}, {Kastner}, {Li}, {Harnden}, {Wolk},
  {Murray}, {Lada}, {Muench}, {McCaughrean}, {Meeus}, {Damiani}, {Micela},
  {Sciortino}, {Bally}, {Hillenbrand}, {Herbst}, {Preibisch}, \&
  {Feigelson}}]{2005ApJSGetmanA}
{Getman}, K.~V., {Flaccomio}, E., {Broos}, P.~S., {et~al.} 2005, \apjs, 160,
  319

\bibitem[{{G{\'o}mez de Castro}(2009)}]{2009ApSSGomezDeCastro}
{G{\'o}mez de Castro}, A.~I. 2009, \apss, 320, 97

\bibitem[{{Gorti} {et~al.}(2009){Gorti}, {Dullemond}, \&
  {Hollenbach}}]{2009ApJGortiB}
{Gorti}, U., {Dullemond}, C.~P., \& {Hollenbach}, D. 2009, \apj, 705, 1237

\bibitem[{{G{\"u}del} {et~al.}(2007){G{\"u}del}, {Briggs}, {Arzner}, {Audard},
  {Bouvier}, {Feigelson}, {Franciosini}, {Glauser}, {Grosso}, {Micela},
  {Monin}, {Montmerle}, {Padgett}, {Palla}, {Pillitteri}, {Rebull}, {Scelsi},
  {Silva}, {Skinner}, {Stelzer}, \& {Telleschi}}]{2007AAGudel}
{G{\"u}del}, M., {Briggs}, K.~R., {Arzner}, K., {et~al.} 2007, \aap, 468, 353

\bibitem[{{Gundlach} {et~al.}(2018){Gundlach}, {Schmidt}, {Kreuzig},
  {Bischoff}, {Rezaei}, {Kothe}, {Blum}, {Grzesik}, \&
  {Stoll}}]{2018MNRASGundlach}
{Gundlach}, B., {Schmidt}, K.~P., {Kreuzig}, C., {et~al.} 2018, \mnras, 479,
  1273

\bibitem[{{Habing}(1968)}]{1968BANHabing}
{Habing}, H.~J. 1968, \bain, 19, 421

\bibitem[{{Haisch} {et~al.}(2001){Haisch}, {Lada}, \& {Lada}}]{2001ApJHaisch}
{Haisch}, Karl~E., J., {Lada}, E.~A., \& {Lada}, C.~J. 2001, \apjl, 553, L153

\bibitem[{{Hartmann} {et~al.}(2016){Hartmann}, {Herczeg}, \&
  {Calvet}}]{2016ARAAHartmann}
{Hartmann}, L., {Herczeg}, G., \& {Calvet}, N. 2016, \araa, 54, 135

\bibitem[{{Haworth} {et~al.}(2018){Haworth}, {Clarke}, {Rahman}, {Winter}, \&
  {Facchini}}]{2018MNRASHaworth}
{Haworth}, T.~J., {Clarke}, C.~J., {Rahman}, W., {Winter}, A.~J., \&
  {Facchini}, S. 2018, \mnras, 481, 452

\bibitem[{{Hendler} {et~al.}(2020){Hendler}, {Pascucci}, {Pinilla}, {Tazzari},
  {Carpenter}, {Malhotra}, \& {Testi}}]{2020ApJHendler}
{Hendler}, N., {Pascucci}, I., {Pinilla}, P., {et~al.} 2020, \apj, 895, 126

\bibitem[{{Hennebelle} {et~al.}(2016){Hennebelle}, {Commer{\c{c}}on},
  {Chabrier}, \& {Marchand}}]{2016ApJHennebelle}
{Hennebelle}, P., {Commer{\c{c}}on}, B., {Chabrier}, G., \& {Marchand}, P.
  2016, \apjl, 830, L8

\bibitem[{{Hollenbach} {et~al.}(1994){Hollenbach}, {Johnstone}, {Lizano}, \&
  {Shu}}]{1994ApJHollenbach}
{Hollenbach}, D., {Johnstone}, D., {Lizano}, S., \& {Shu}, F. 1994, \apj, 428,
  654

\bibitem[{{Huang} {et~al.}(2020){Huang}, {Andrews}, {{\"O}berg}, {Ansdell},
  {Benisty}, {Carpenter}, {Isella}, {P{\'e}rez}, {Ricci}, {Williams}, {Wilner},
  \& {Zhu}}]{2020ApJHuang}
{Huang}, J., {Andrews}, S.~M., {{\"O}berg}, K.~I., {et~al.} 2020, \apj, 898,
  140

\bibitem[{{Hueso} \& {Guillot}(2005)}]{2005A&AHueso}
{Hueso}, R. \& {Guillot}, T. 2005, \aap, 442, 703

\bibitem[{{Hunter}(2007)}]{2007CSEHunter}
{Hunter}, J.~D. 2007, Computing in Science and Engineering, 9, 90

\bibitem[{{Johnstone} {et~al.}(2021){Johnstone}, {Bartel}, \&
  {G{\"u}del}}]{2021AAJohnstone}
{Johnstone}, C.~P., {Bartel}, M., \& {G{\"u}del}, M. 2021, \aap, 649, A96

\bibitem[{{Kennedy} \& {Kenyon}(2009)}]{2009ApJKennedyKenyon}
{Kennedy}, G.~M. \& {Kenyon}, S.~J. 2009, \apj, 695, 1210

\bibitem[{{Kimura} {et~al.}(2016){Kimura}, {Kunitomo}, \&
  {Takahashi}}]{2016MNRASKimura}
{Kimura}, S.~S., {Kunitomo}, M., \& {Takahashi}, S.~Z. 2016, \mnras, 461, 2257

\bibitem[{{Kingma} \& {Ba}(2014)}]{adam}
{Kingma}, D.~P. \& {Ba}, J. 2014, arXiv e-prints, arXiv:1412.6980

\bibitem[{{Koepferl} {et~al.}(2013){Koepferl}, {Ercolano}, {Dale}, {Teixeira},
  {Ratzka}, \& {Spezzi}}]{2013MNRASKoepferl}
{Koepferl}, C.~M., {Ercolano}, B., {Dale}, J., {et~al.} 2013, \mnras, 428, 3327

\bibitem[{{Komaki} {et~al.}(2021){Komaki}, {Nakatani}, \&
  {Yoshida}}]{2021ApJKomaki}
{Komaki}, A., {Nakatani}, R., \& {Yoshida}, N. 2021, \apj, 910, 51

\bibitem[{{Kraus} {et~al.}(2012){Kraus}, {Ireland}, {Hillenbrand}, \&
  {Martinache}}]{2012ApJKraus}
{Kraus}, A.~L., {Ireland}, M.~J., {Hillenbrand}, L.~A., \& {Martinache}, F.
  2012, \apj, 745, 19

\bibitem[{{Kuffmeier} {et~al.}(2020){Kuffmeier}, {Goicovic}, \&
  {Dullemond}}]{2020AAKuffmeierA}
{Kuffmeier}, M., {Goicovic}, F.~G., \& {Dullemond}, C.~P. 2020, \aap, 633, A3

\bibitem[{{Kunitomo} {et~al.}(2021){Kunitomo}, {Ida}, {Takeuchi}, {Pani{\'c}},
  {Miley}, \& {Suzuki}}]{2021ApJKunitomo}
{Kunitomo}, M., {Ida}, S., {Takeuchi}, T., {et~al.} 2021, \apj, 909, 109

\bibitem[{{Kunitomo} {et~al.}(2020){Kunitomo}, {Suzuki}, \&
  {Inutsuka}}]{2020MNRASKunitomo}
{Kunitomo}, M., {Suzuki}, T.~K., \& {Inutsuka}, S.-i. 2020, \mnras, 492, 3849

\bibitem[{{Lada} \& {Lada}(2003)}]{2003ARAALadaLada}
{Lada}, C.~J. \& {Lada}, E.~A. 2003, \araa, 41, 57

\bibitem[{{Lebreuilly} {et~al.}(2021){Lebreuilly}, {Hennebelle}, {Colman},
  {Commer{\c{c}}on}, {Klessen}, {Maury}, {Molinari}, \&
  {Testi}}]{2021ApJLebreuilly}
{Lebreuilly}, U., {Hennebelle}, P., {Colman}, T., {et~al.} 2021, \apjl, 917,
  L10

\bibitem[{{Lee} {et~al.}(2021){Lee}, {Marchand}, {Liu}, \&
  {Hennebelle}}]{2021ApJLee}
{Lee}, Y.-N., {Marchand}, P., {Liu}, Y.-H., \& {Hennebelle}, P. 2021, \apj,
  922, 36

\bibitem[{{Lenz} {et~al.}(2019){Lenz}, {Klahr}, \& {Birnstiel}}]{2019ApJLenz}
{Lenz}, C.~T., {Klahr}, H., \& {Birnstiel}, T. 2019, \apj, 874, 36

\bibitem[{{Lesur} {et~al.}(2014){Lesur}, {Kunz}, \& {Fromang}}]{2014AALesur}
{Lesur}, G., {Kunz}, M.~W., \& {Fromang}, S. 2014, \aap, 566, A56

\bibitem[{{Lodders}(2003)}]{2003ApJLodders}
{Lodders}, K. 2003, \apj, 591, 1220

\bibitem[{{Luhman}(2000)}]{2000ApJLuhman}
{Luhman}, K.~L. 2000, \apj, 544, 1044

\bibitem[{{L{\"u}st}(1952)}]{1952ZNatALust}
{L{\"u}st}, R. 1952, Zeitschrift Naturforschung Teil A, 7, 87

\bibitem[{{Lynden-Bell} \& {Pringle}(1974)}]{1974NMRASLyndenBellPringle}
{Lynden-Bell}, D. \& {Pringle}, J.~E. 1974, \mnras, 168, 603

\bibitem[{{Mamajek}(2009)}]{2009AIPCMamajek}
{Mamajek}, E.~E. 2009, in American Institute of Physics Conference Series, Vol.
  1158, American Institute of Physics Conference Series, ed. T.~{Usuda},
  M.~{Tamura}, \& M.~{Ishii}, 3--10

\bibitem[{{Manara} {et~al.}(2016{\natexlab{a}}){Manara}, {Fedele}, {Herczeg},
  \& {Teixeira}}]{2016AAManaraA}
{Manara}, C.~F., {Fedele}, D., {Herczeg}, G.~J., \& {Teixeira}, P.~S.
  2016{\natexlab{a}}, \aap, 585, A136

\bibitem[{{Manara} {et~al.}(2019){Manara}, {Mordasini}, {Testi}, {Williams},
  {Miotello}, {Lodato}, \& {Emsenhuber}}]{2019AAManara}
{Manara}, C.~F., {Mordasini}, C., {Testi}, L., {et~al.} 2019, \aap, 631, L2

\bibitem[{{Manara} {et~al.}(2012){Manara}, {Robberto}, {Da Rio}, {Lodato},
  {Hillenbrand}, {Stassun}, \& {Soderblom}}]{2012ApJManara}
{Manara}, C.~F., {Robberto}, M., {Da Rio}, N., {et~al.} 2012, \apj, 755, 154

\bibitem[{{Manara} {et~al.}(2016{\natexlab{b}}){Manara}, {Rosotti}, {Testi},
  {Natta}, {Alcal{\'a}}, {Williams}, {Ansdell}, {Miotello}, {van der Marel},
  {Tazzari}, {Carpenter}, {Guidi}, {Mathews}, {Oliveira}, {Prusti}, \& {van
  Dishoeck}}]{2016AAManaraB}
{Manara}, C.~F., {Rosotti}, G., {Testi}, L., {et~al.} 2016{\natexlab{b}}, \aap,
  591, L3

\bibitem[{{Manara} {et~al.}(2017){Manara}, {Testi}, {Herczeg}, {Pascucci},
  {Alcal{\'a}}, {Natta}, {Antoniucci}, {Fedele}, {Mulders}, {Henning},
  {Mohanty}, {Prusti}, \& {Rigliaco}}]{2017AAManara}
{Manara}, C.~F., {Testi}, L., {Herczeg}, G.~J., {et~al.} 2017, \aap, 604, A127

\bibitem[{{Matsuyama} {et~al.}(2003){Matsuyama}, {Johnstone}, \&
  {Hartmann}}]{2003ApJMatsuyama}
{Matsuyama}, I., {Johnstone}, D., \& {Hartmann}, L. 2003, \apj, 582, 893

\bibitem[{McKay {et~al.}(1979)McKay, Beckman, \& Conover}]{1979TechMcKay}
McKay, M.~D., Beckman, R.~J., \& Conover, W.~J. 1979, Technometrics, 21, 239

\bibitem[{{Michel} {et~al.}(2021){Michel}, {van der Marel}, \&
  {Matthews}}]{2021ApJMichel}
{Michel}, A., {van der Marel}, N., \& {Matthews}, B.~C. 2021, \apj, 921, 72

\bibitem[{{Mordasini} {et~al.}(2009){Mordasini}, {Alibert}, \&
  {Benz}}]{2009A&AMordasinia}
{Mordasini}, C., {Alibert}, Y., \& {Benz}, W. 2009, \aap, 501, 1139

\bibitem[{{Mulders} {et~al.}(2017){Mulders}, {Pascucci}, {Manara}, {Testi},
  {Herczeg}, {Henning}, {Mohanty}, \& {Lodato}}]{2017ApJMulders}
{Mulders}, G.~D., {Pascucci}, I., {Manara}, C.~F., {et~al.} 2017, \apj, 847, 31

\bibitem[{{Murray} {et~al.}(2001){Murray}, {Chaboyer}, {Arras}, {Hansen}, \&
  {Noyes}}]{2001ApJMurray}
{Murray}, N., {Chaboyer}, B., {Arras}, P., {Hansen}, B., \& {Noyes}, R.~W.
  2001, \apj, 555, 801

\bibitem[{{Musiolik} \& {Wurm}(2019)}]{2019ApJMusiolikWurm}
{Musiolik}, G. \& {Wurm}, G. 2019, \apj, 873, 58

\bibitem[{{Nakagawa} {et~al.}(1986){Nakagawa}, {Sekiya}, \&
  {Hayashi}}]{1986IcarusNakagawa}
{Nakagawa}, Y., {Sekiya}, M., \& {Hayashi}, C. 1986, \icarus, 67, 375

\bibitem[{{Nakamoto} \& {Nakagawa}(1994)}]{1994ApJNakamoto}
{Nakamoto}, T. \& {Nakagawa}, Y. 1994, \apj, 421, 640

\bibitem[{{Nayakshin} {et~al.}(2022){Nayakshin}, {Elbakyan}, \&
  {Rosotti}}]{2022MNRASNayakshin}
{Nayakshin}, S., {Elbakyan}, V., \& {Rosotti}, G. 2022, \mnras, 512, 6038

\bibitem[{{Nayakshin} {et~al.}(2020){Nayakshin}, {Tsukagoshi}, {Hall}, {Vazan},
  {Helled}, {Humphries}, {Meru}, {Neunteufel}, \&
  {Panic}}]{2020MNRASNayakshinB}
{Nayakshin}, S., {Tsukagoshi}, T., {Hall}, C., {et~al.} 2020, \mnras, 495, 285

\bibitem[{{Owen} {et~al.}(2012){Owen}, {Clarke}, \& {Ercolano}}]{2012MNRASOwen}
{Owen}, J.~E., {Clarke}, C.~J., \& {Ercolano}, B. 2012, \mnras, 422, 1880

\bibitem[{{Owen} {et~al.}(2011){Owen}, {Ercolano}, \& {Clarke}}]{2011MNRASOwen}
{Owen}, J.~E., {Ercolano}, B., \& {Clarke}, C.~J. 2011, \mnras, 412, 13

\bibitem[{{Pascucci} {et~al.}(2022){Pascucci}, {Cabrit}, {Edwards}, {Gorti},
  {Gressel}, \& {Suzuki}}]{2022PPVIIPascucci}
{Pascucci}, I., {Cabrit}, S., {Edwards}, S., {et~al.} 2022, arXiv e-prints,
  arXiv:2203.10068

\bibitem[{{Pascucci} {et~al.}(2016){Pascucci}, {Testi}, {Herczeg}, {Long},
  {Manara}, {Hendler}, {Mulders}, {Krijt}, {Ciesla}, {Henning}, {Mohanty},
  {Drabek-Maunder}, {Apai}, {Sz{\H{u}}cs}, {Sacco}, \&
  {Olofsson}}]{2016ApJPascucci}
{Pascucci}, I., {Testi}, L., {Herczeg}, G.~J., {et~al.} 2016, \apj, 831, 125

\bibitem[{Pedregosa {et~al.}(2011)Pedregosa, Varoquaux, Gramfort, Michel,
  Thirion, Grisel, Blondel, Prettenhofer, Weiss, Dubourg, Vanderplas, Passos,
  Cournapeau, Brucher, Perrot, \& Duchesnay}]{sklearn}
Pedregosa, F., Varoquaux, G., Gramfort, A., {et~al.} 2011, Journal of Machine
  Learning Research, 12, 2825

\bibitem[{{Pfalzner} {et~al.}(2022){Pfalzner}, {Dehghani}, \&
  {Michel}}]{2022ApJPfalzner}
{Pfalzner}, S., {Dehghani}, S., \& {Michel}, A. 2022, \apjl, 939, L10

\bibitem[{{Picogna} {et~al.}(2021){Picogna}, {Ercolano}, \&
  {Espaillat}}]{2021MNRASPicogna}
{Picogna}, G., {Ercolano}, B., \& {Espaillat}, C.~C. 2021, \mnras, 508, 3611

\bibitem[{{Picogna} {et~al.}(2019){Picogna}, {Ercolano}, {Owen}, \&
  {Weber}}]{2019MNRASPicogna}
{Picogna}, G., {Ercolano}, B., {Owen}, J.~E., \& {Weber}, M.~L. 2019, \mnras,
  487, 691

\bibitem[{{Powell} {et~al.}(2019){Powell}, {Murray-Clay}, {P{\'e}rez},
  {Schlichting}, \& {Rosenthal}}]{2019ApJPowell}
{Powell}, D., {Murray-Clay}, R., {P{\'e}rez}, L.~M., {Schlichting}, H.~E., \&
  {Rosenthal}, M. 2019, \apj, 878, 116

\bibitem[{{Preibisch} {et~al.}(2005){Preibisch}, {Kim}, {Favata}, {Feigelson},
  {Flaccomio}, {Getman}, {Micela}, {Sciortino}, {Stassun}, {Stelzer}, \&
  {Zinnecker}}]{2005ApJSPreibischA}
{Preibisch}, T., {Kim}, Y.-C., {Favata}, F., {et~al.} 2005, \apjs, 160, 401

\bibitem[{{Preibisch} {et~al.}(1996){Preibisch}, {Zinnecker}, \&
  {Herbig}}]{1996AAPreibisch}
{Preibisch}, T., {Zinnecker}, H., \& {Herbig}, G.~H. 1996, \aap, 310, 456

\bibitem[{{Qiao} {et~al.}(2022){Qiao}, {Haworth}, {Sellek}, \&
  {Ali}}]{2022MNRASQiao}
{Qiao}, L., {Haworth}, T.~J., {Sellek}, A.~D., \& {Ali}, A.~A. 2022, \mnras,
  512, 3788

\bibitem[{{Ribas} {et~al.}(2015){Ribas}, {Bouy}, \& {Mer{\'\i}n}}]{2015AARibas}
{Ribas}, {\'A}., {Bouy}, H., \& {Mer{\'\i}n}, B. 2015, \aap, 576, A52

\bibitem[{{Ribas} {et~al.}(2014){Ribas}, {Mer{\'\i}n}, {Bouy}, \&
  {Maud}}]{2014AARibas}
{Ribas}, {\'A}., {Mer{\'\i}n}, B., {Bouy}, H., \& {Maud}, L.~T. 2014, \aap,
  561, A54

\bibitem[{{Rosotti} {et~al.}(2017){Rosotti}, {Clarke}, {Manara}, \&
  {Facchini}}]{2017MNRASRosotti}
{Rosotti}, G.~P., {Clarke}, C.~J., {Manara}, C.~F., \& {Facchini}, S. 2017,
  \mnras, 468, 1631

\bibitem[{{Santos} {et~al.}(2005){Santos}, {Israelian}, {Mayor}, {Bento},
  {Almeida}, {Sousa}, \& {Ecuvillon}}]{2005AASantos}
{Santos}, N.~C., {Israelian}, G., {Mayor}, M., {et~al.} 2005, \aap, 437, 1127

\bibitem[{{Sellek} {et~al.}(2020){Sellek}, {Booth}, \&
  {Clarke}}]{2020MNRASSellek}
{Sellek}, A.~D., {Booth}, R.~A., \& {Clarke}, C.~J. 2020, \mnras, 498, 2845

\bibitem[{{Shakura} \& {Sunyaev}(1973)}]{1973A&AShakuraSunyaev}
{Shakura}, N.~I. \& {Sunyaev}, R.~A. 1973, \aap, 500, 33

\bibitem[{{Somigliana} {et~al.}(2020){Somigliana}, {Toci}, {Lodato}, {Rosotti},
  \& {Manara}}]{2020MNRASSomigliana}
{Somigliana}, A., {Toci}, C., {Lodato}, G., {Rosotti}, G., \& {Manara}, C.~F.
  2020, \mnras, 492, 1120

\bibitem[{{Somigliana} {et~al.}(2022){Somigliana}, {Toci}, {Rosotti}, {Lodato},
  {Tazzari}, {Manara}, {Testi}, \& {Lepri}}]{2022MNRASSomigliana}
{Somigliana}, A., {Toci}, C., {Rosotti}, G., {et~al.} 2022, \mnras, 514, 5927

\bibitem[{{Steinpilz} {et~al.}(2019){Steinpilz}, {Teiser}, \&
  {Wurm}}]{2019ApJSteinpilz}
{Steinpilz}, T., {Teiser}, J., \& {Wurm}, G. 2019, \apj, 874, 60

\bibitem[{{Suzuki} \& {Inutsuka}(2009)}]{2009ApJSuzukiInutsuka}
{Suzuki}, T.~K. \& {Inutsuka}, S.-i. 2009, \apjl, 691, L49

\bibitem[{{Suzuki} {et~al.}(2010){Suzuki}, {Muto}, \&
  {Inutsuka}}]{2010ApJSuzuki}
{Suzuki}, T.~K., {Muto}, T., \& {Inutsuka}, S.-i. 2010, \apj, 718, 1289

\bibitem[{{Suzuki} {et~al.}(2016){Suzuki}, {Ogihara}, {Morbidelli}, {Crida}, \&
  {Guillot}}]{2016AASuzuki}
{Suzuki}, T.~K., {Ogihara}, M., {Morbidelli}, A., {Crida}, A., \& {Guillot}, T.
  2016, \aap, 596, A74

\bibitem[{{Tabone} {et~al.}(2022){Tabone}, {Rosotti}, {Cridland}, {Armitage},
  \& {Lodato}}]{2022MNRASTaboneA}
{Tabone}, B., {Rosotti}, G.~P., {Cridland}, A.~J., {Armitage}, P.~J., \&
  {Lodato}, G. 2022, \mnras, 512, 2290

\bibitem[{{Testi} {et~al.}(2022){Testi}, {Natta}, {Manara}, {de Gregorio
  Monsalvo}, {Lodato}, {Lopez}, {Muzic}, {Pascucci}, {Sanchis}, {Miranda},
  {Scholz}, {De Simone}, \& {Williams}}]{2022AATesti}
{Testi}, L., {Natta}, A., {Manara}, C.~F., {et~al.} 2022, \aap, 663, A98

\bibitem[{{Throop} \& {Bally}(2008)}]{2008AJThroopBally}
{Throop}, H.~B. \& {Bally}, J. 2008, \aj, 135, 2380

\bibitem[{{Tobin} {et~al.}(2020){Tobin}, {Sheehan}, {Megeath},
  {D{\'\i}az-Rodr{\'\i}guez}, {Offner}, {Murillo}, {van 't Hoff}, {van
  Dishoeck}, {Osorio}, {Anglada}, {Furlan}, {Stutz}, {Reynolds}, {Karnath},
  {Fischer}, {Persson}, {Looney}, {Li}, {Stephens}, {Chandler}, {Cox},
  {Dunham}, {Tychoniec}, {Kama}, {Kratter}, {Kounkel}, {Mazur}, {Maud},
  {Patel}, {Perez}, {Sadavoy}, {Segura-Cox}, {Sharma}, {Stephenson}, {Watson},
  \& {Wyrowski}}]{2020ApJTobinA}
{Tobin}, J.~J., {Sheehan}, P.~D., {Megeath}, S.~T., {et~al.} 2020, \apj, 890,
  130

\bibitem[{{Toci} {et~al.}(2021){Toci}, {Rosotti}, {Lodato}, {Testi}, \&
  {Trapman}}]{2021MNRASToci}
{Toci}, C., {Rosotti}, G., {Lodato}, G., {Testi}, L., \& {Trapman}, L. 2021,
  \mnras, 507, 818

\bibitem[{{Tripathi} {et~al.}(2017){Tripathi}, {Andrews}, {Birnstiel}, \&
  {Wilner}}]{2017ApJTripathi}
{Tripathi}, A., {Andrews}, S.~M., {Birnstiel}, T., \& {Wilner}, D.~J. 2017,
  \apj, 845, 44

\bibitem[{{Tychoniec} {et~al.}(2020){Tychoniec}, {Manara}, {Rosotti}, {van
  Dishoeck}, {Cridland}, {Hsieh}, {Murillo}, {Segura-Cox}, {van Terwisga}, \&
  {Tobin}}]{2020AATychoniec}
{Tychoniec}, {\L}., {Manara}, C.~F., {Rosotti}, G.~P., {et~al.} 2020, \aap,
  640, A19

\bibitem[{{Tychoniec} {et~al.}(2018){Tychoniec}, {Tobin}, {Karska}, {Chandler},
  {Dunham}, {Harris}, {Kratter}, {Li}, {Looney}, \&
  {Melis}}]{2018ApJSTychoniec}
{Tychoniec}, {\L}., {Tobin}, J.~J., {Karska}, A., {et~al.} 2018, \apjs, 238, 19

\bibitem[{{van Terwisga} {et~al.}(2020){van Terwisga}, {van Dishoeck}, {Mann},
  {Di Francesco}, {van der Marel}, {Meyer}, {Andrews}, {Carpenter}, {Eisner},
  {Manara}, \& {Williams}}]{2020AAvanTerwisga}
{van Terwisga}, S.~E., {van Dishoeck}, E.~F., {Mann}, R.~K., {et~al.} 2020,
  \aap, 640, A27

\bibitem[{{Venuti} {et~al.}(2017){Venuti}, {Bouvier}, {Cody}, {Stauffer},
  {Micela}, {Rebull}, {Alencar}, {Sousa}, {Hillenbrand}, \&
  {Flaccomio}}]{2017AAVenuti}
{Venuti}, L., {Bouvier}, J., {Cody}, A.~M., {et~al.} 2017, \aap, 599, A23

\bibitem[{{Veras} \& {Armitage}(2004)}]{2004MNRASVerasArmitage}
{Veras}, D. \& {Armitage}, P.~J. 2004, \mnras, 347, 613

\bibitem[{Virtanen {et~al.}(2020)Virtanen, Gommers, Oliphant, Haberland, Reddy,
  Cournapeau, Burovski, Peterson, Weckesser, Bright, {van der Walt}, Brett,
  Wilson, Millman, Mayorov, Nelson, Jones, Kern, Larson, Carey, Polat, Feng,
  Moore, {VanderPlas}, Laxalde, Perktold, Cimrman, Henriksen, Quintero, Harris,
  Archibald, Ribeiro, Pedregosa, {van Mulbregt}, \& {SciPy 1.0
  Contributors}}]{2020NatMethScipy}
Virtanen, P., Gommers, R., Oliphant, T.~E., {et~al.} 2020, Nature Methods, 17,
  261

\bibitem[{{Voelkel} {et~al.}(2020){Voelkel}, {Klahr}, {Mordasini},
  {Emsenhuber}, \& {Lenz}}]{2020AAVoelkel}
{Voelkel}, O., {Klahr}, H., {Mordasini}, C., {Emsenhuber}, A., \& {Lenz}, C.
  2020, \aap, 642, A75

\bibitem[{{Weder} {et~al.}(in prep.){Weder}, {Mordasini}, \&
  {Emsenhuber}}]{Weder2023}
{Weder}, J., {Mordasini}, C., \& {Emsenhuber}, A. in prep., \aap

\bibitem[{{Williams} {et~al.}(2019){Williams}, {Cieza}, {Hales}, {Ansdell},
  {Ruiz-Rodriguez}, {Casassus}, {Perez}, \& {Zurlo}}]{2019ApJWilliams}
{Williams}, J.~P., {Cieza}, L., {Hales}, A., {et~al.} 2019, \apjl, 875, L9

\bibitem[{{Winter} \& {Haworth}(2022)}]{2022EPJPWinterHaworth}
{Winter}, A.~J. \& {Haworth}, T.~J. 2022, European Physical Journal Plus, 137,
  1132

\bibitem[{{Zagaria} {et~al.}(2022){Zagaria}, {Clarke}, {Rosotti}, \&
  {Manara}}]{2022MNRASZagariaA}
{Zagaria}, F., {Clarke}, C.~J., {Rosotti}, G.~P., \& {Manara}, C.~F. 2022,
  \mnras, 512, 3538

\bibitem[{{Zormpas} {et~al.}(2022){Zormpas}, {Birnstiel}, {Rosotti}, \&
  {Andrews}}]{2022AAZormpas}
{Zormpas}, A., {Birnstiel}, T., {Rosotti}, G.~P., \& {Andrews}, S.~M. 2022,
  \aap, 661, A66

\end{thebibliography}

\begin{appendix}

\section{Parameter study}
\label{sec:paramstudy}

\begin{figure*}
	\centering
	\includegraphics{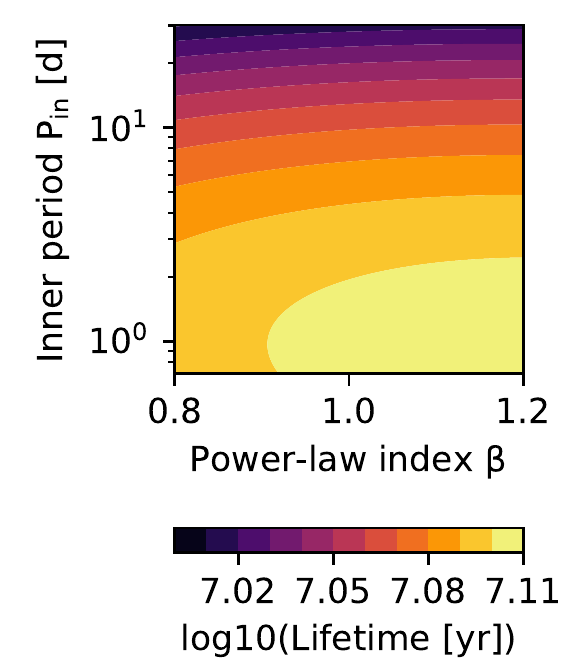}
	\includegraphics{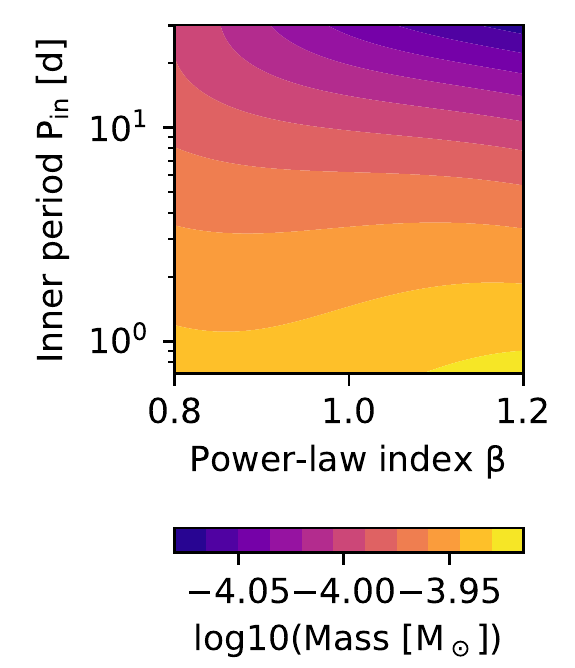}
	\includegraphics{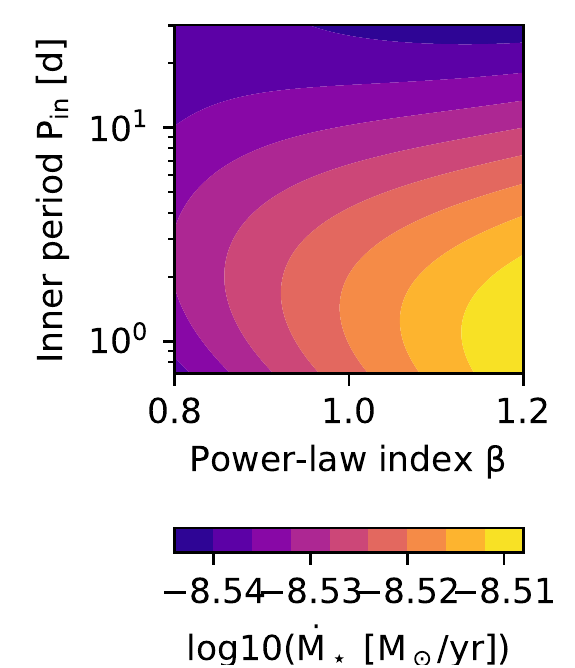}
	\includegraphics{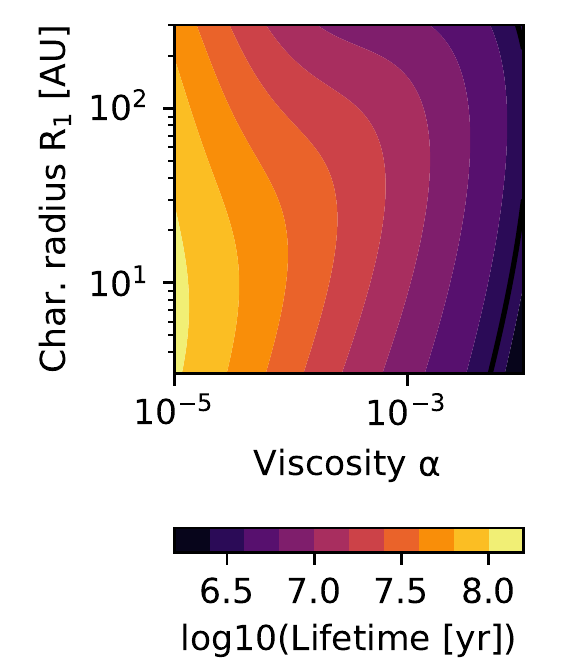}
	\includegraphics{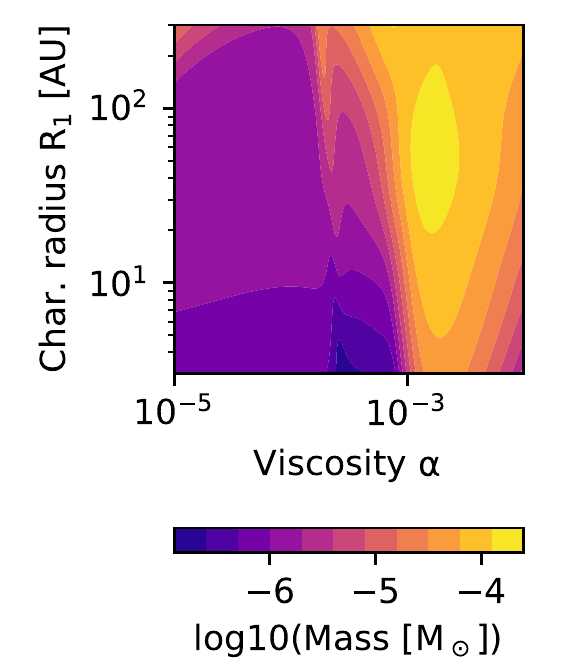}
	\includegraphics{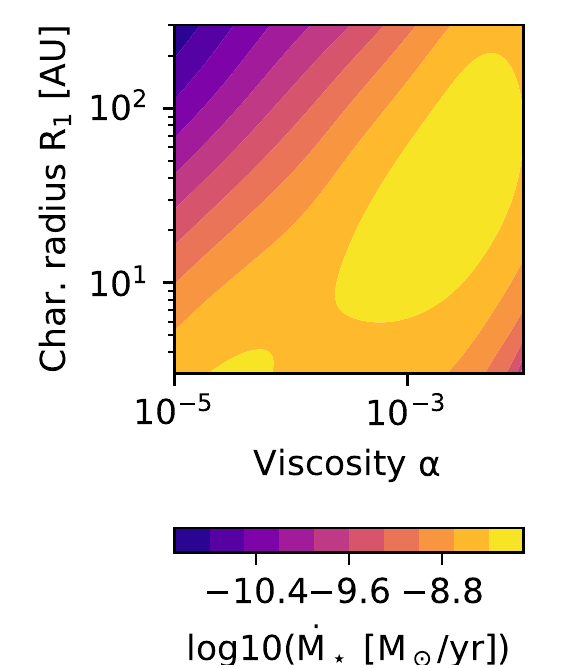}
	\caption{Gas disc lifetime (left), observed dust masses at \SI{2}{\mega\year} (centre), and stellar accretion rates at \SI{2}{\mega\year} (right) as functions of the power-law index of the initial profile $\beta$ and the inner edge $r_\mathrm{in}$ (top) or viscosity parameter $\alpha$ and characteristic radius $r_1$ (bottom) In the bottom row, the white region is where disc lifetimes are less than \SI{2}{\mega\year} and the values cannot be constrained by the model.}
	\label{fig:param-study-1}
\end{figure*}

\begin{figure*}
	\centering
	\includegraphics{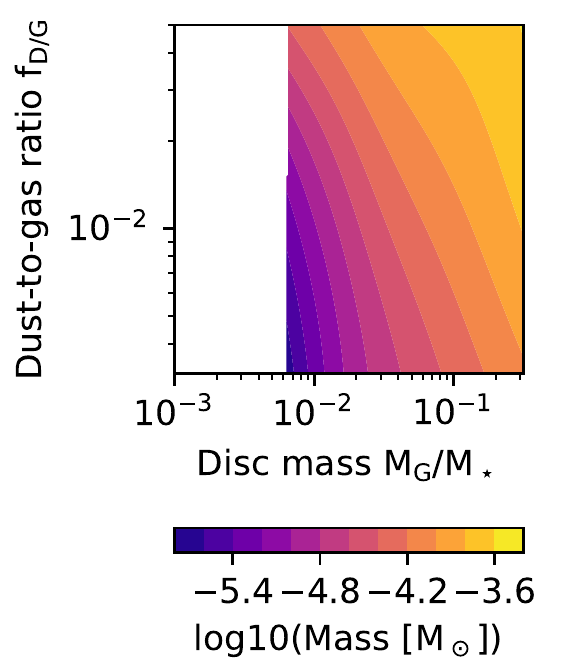}
	\includegraphics{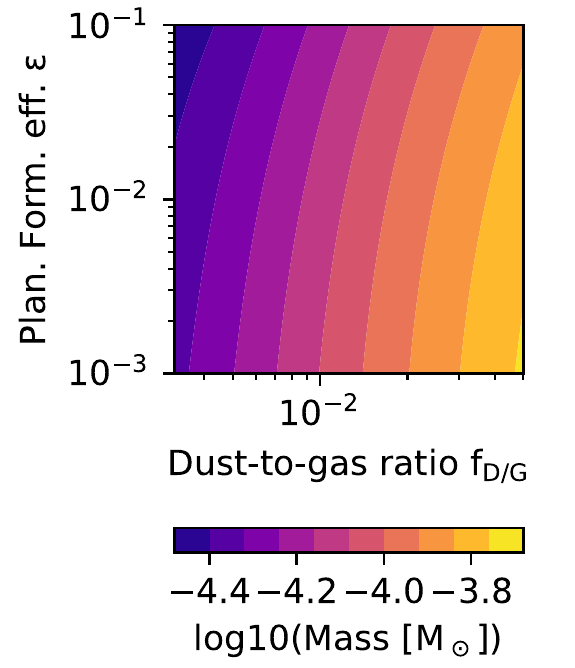}
	\includegraphics{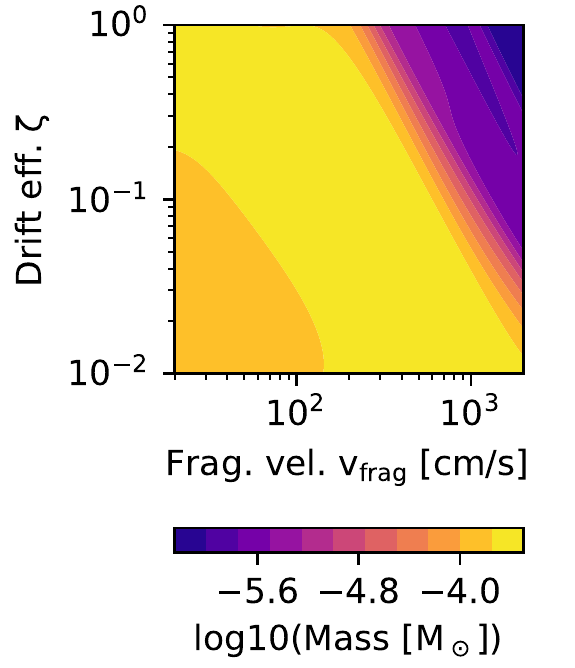}
	\caption{Observed mass at \SI{2}{\mega\year} as function of the parameters of the \textit{twopop} model: gas-disc mass and dust-to-gas ratio (left), planetesimal formation efficiency and fragmentation velocity (centre), and drift efficiency and fragmentation velocity (right).}
	\label{fig:param-study-2}
\end{figure*}

The surrogate models allow us to study the effects of the different parameters, as shown for stellar mass and photoevaporation in Sect.~\ref{sec:res-photo}. Here, we expand the study to other parameters. The parameters that are not varied are selected as in Sect.~\ref{sec:res-photo}, except for the photoevaporation-related parameters, which are set as $L_\mathrm{X}=\SI{1e29}{\erg\per\second}$ and $F=\SI{10}{G_0}$ to provide lifetimes that are globally in line with the observations.

We study in Fig.~\ref{fig:param-study-1} the effects of the parameters of the gas disc model. The top row shows the outcomes as functions of the power-law index of the initial profile $\beta$ and the inner edge $r_\mathrm{in}$. These two parameters have very little effect on the final lifetimes, as all values lie within about \SI{0.1}{dex}, corresponding to a maximum relative difference of \SI{26}{\percent}. The same applies to disc masses and stellar accretion rates. Thus, the choice of these parameters has negligible effects on the final properties and we do not discuss these parameters further in this work.

The bottom row of Fig.~\ref{fig:param-study-1} shows the effect of the viscosity parameter $\alpha$ and the disc's characteristic radius $r_1$. The characteristic radius has limited effect on the disc lifetimes while $\alpha$, which affects the whole viscous evolution, has an important effect. However, both parameters affect the stellar accretion rate, making it possible to combine these two parameters to set the behaviour of stellar accretion versus lifetime.

Figure~\ref{fig:param-study-2} shows the same analysis but for the parameters of the solid disc. As such, only the observed dust masses are shown for each parameter combination. The left panel features the dust-to-gas ratio $f_\mathrm{D/G}$ and the disc's gas mass $M_\mathrm{G}/M_\star$. The results show that the dust-to-gas ratio is only important to control the observed disc masses for discs that are close to dispersal (towards the left of the panel) while it has a lower effect on more massive discs. The centre panel shows two of the \textit{twopop} model parameters, planetesimal formation efficiency and dust-to-gas ratio. The panel shows that the planetesimal formation efficiency parameter $\varepsilon$ is of lower importance than the initial dust mass (which is controlled by the dust-to-gas ratio $f_\mathrm{D/G}$) for the observed disc masses. The right panels show two other \textit{twopop} model parameters, the drift efficiency $\zeta$ and the fragmentation velocity $v_\mathrm{frag}$. Here we see that fragmentation velocities $v_\mathrm{frag}\gtrsim\SI{300}{\centi\meter\per\second}$ lead to lower disc masses because drift becomes efficient. This effect can be counterbalanced by reducing the drift efficiency $\zeta$. However, for a value of $v_\mathrm{frag}=\SI{200}{\centi\meter\per\second}$, the drift efficiency has a small effect on the observed disc masses, which indicate that drift is already inefficient.

One might expect observed dust masses to be affected equally by the two parameters shown in the left panel (as the initial solid mass is the product of the dust-to-gas ratio and the gas mass $M_\mathrm{G}$). However, our results indicate that the initial gas mass has a greater effect than the dust-to-gas ratio.

\clearpage

\section{Best parameters for population}
\label{sec:popparams}

\begin{figure*}
	\centering
	\includegraphics{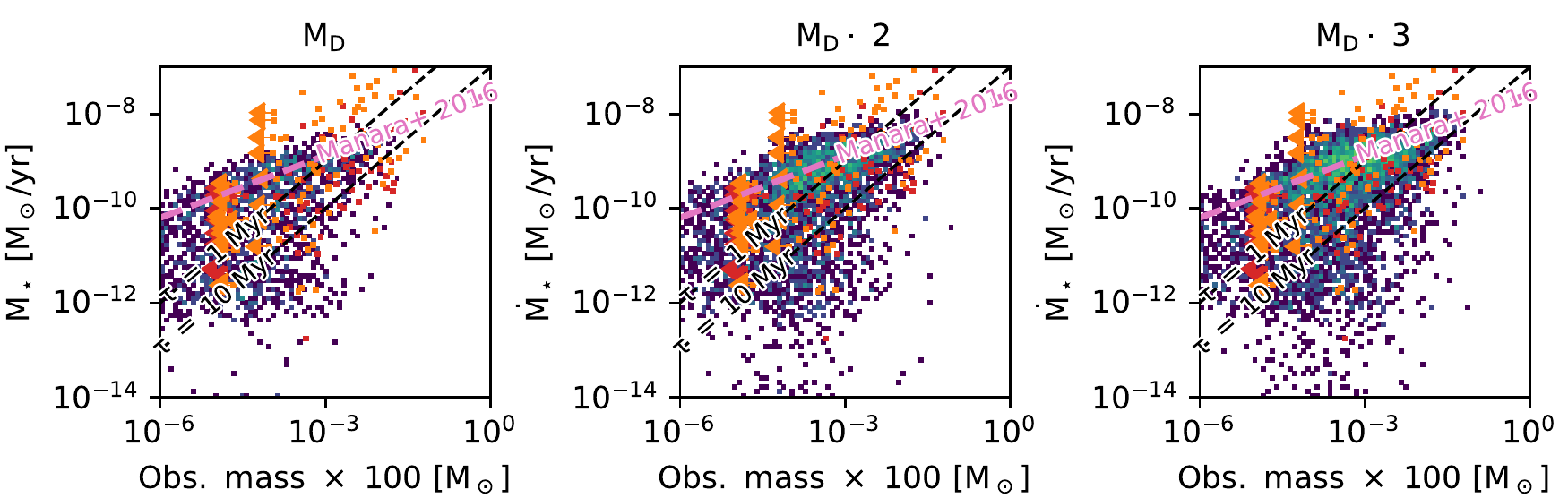}
	\includegraphics{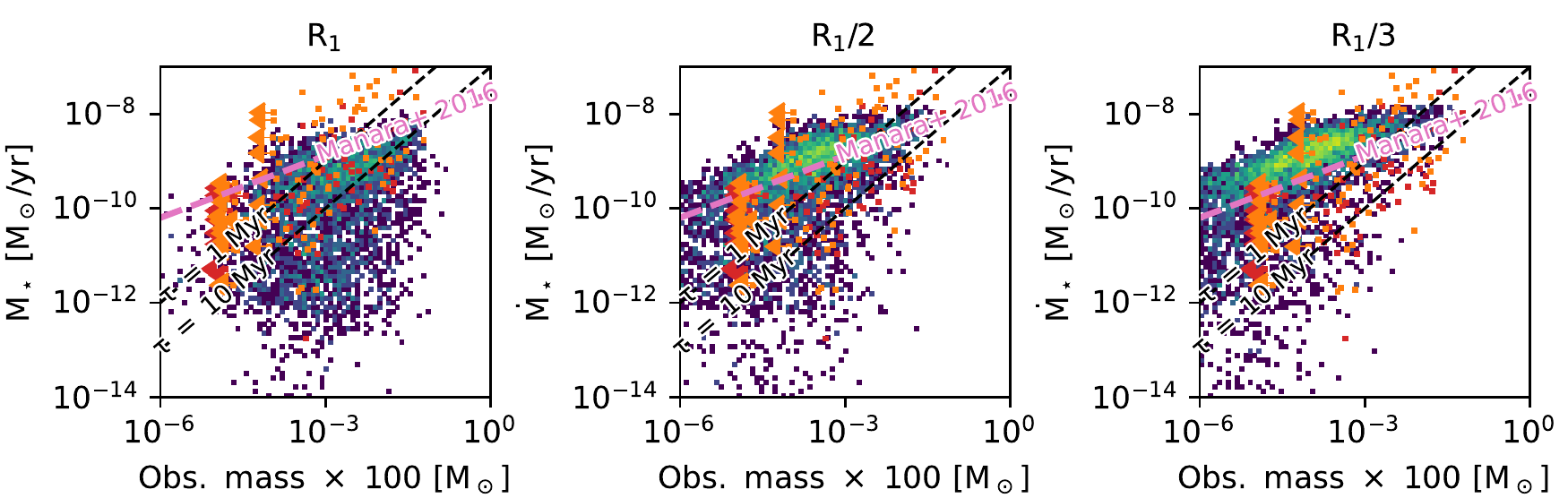}
	\includegraphics{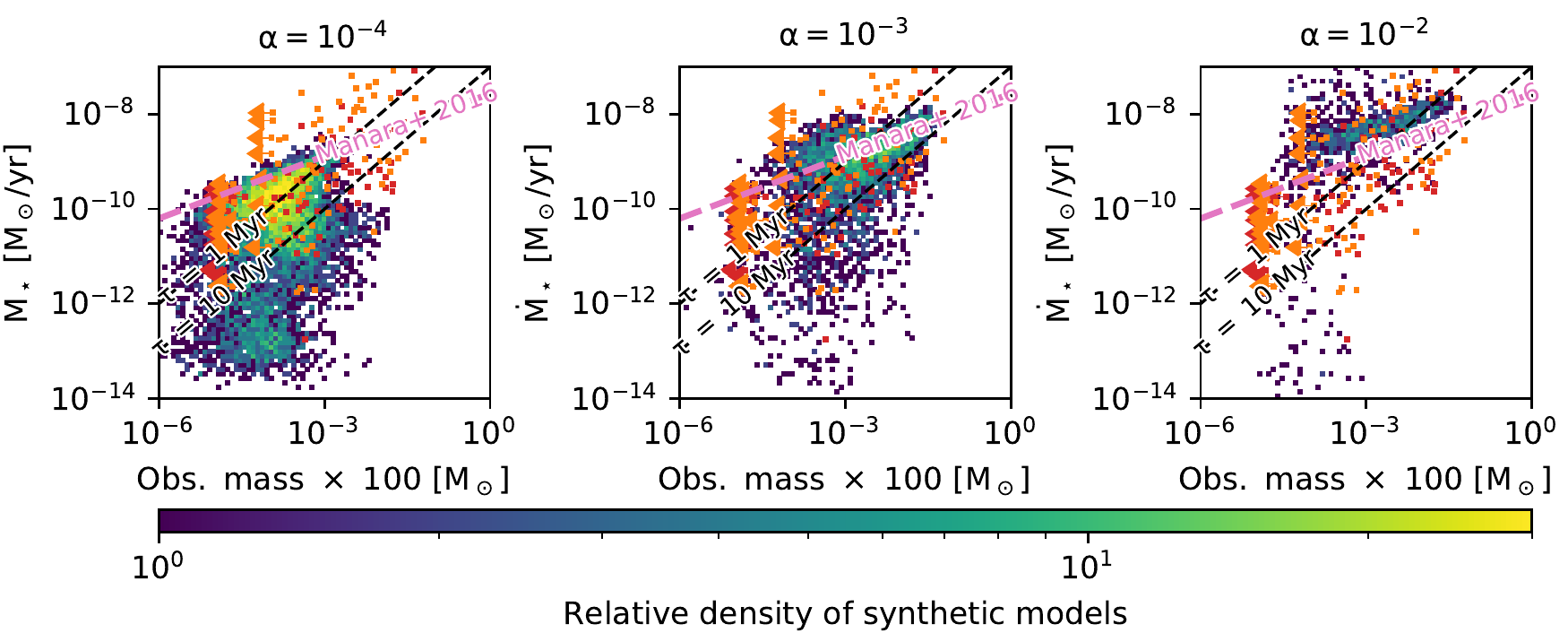}
	\caption{2D histograms for stellar accretion rate versus disc mass at \SI{2}{\mega\year}, showing the effects of the different parameters, as given above each panel. The top row shows the effect of increasing the initial disc mass $M_\mathrm{D}$. The middle row shows the effect of reducing the disc characteristic radius $r_1$. The bottom row shows the effect of the disc viscosity parameter $\alpha$. The observed data from the Lupus (in red) and Chamaeleon~I (orange) star-forming regions from \citet{2019AAManara} are shown for comparison.}
	\label{fig:mar-params}
\end{figure*}

To show how the different model parameters affect the mass--accretion rate relationship and how we came to select the best model parameters, in Fig.~\ref{fig:mar-params}  we provide several 2D histograms similar to those shown in Fig.~\ref{fig:mar-pop-canon}, but varying one parameter at a time. These were generated with the parameters of the `best match' population, except for the parameter being varied.

The first row of the figure shows the effect of varying the disc mass (both gas and solids). We see that an increase in the initial mass is recovered in the observed mass after \SI{2}{\mega\year} of evolution. In addition, the stellar accretion rate is correlated with the disc mass similarly to the best fit of \citet{2016AAManaraB}. The initial disc mass can therefore be used to set observed masses and an increase from the canonical value is required to obtain disc masses consistent with observations. However, this parameter cannot be used to control the behaviour of the stellar accretion rates for given disc masses.

The second row of the same figure shows the effect of reducing the disc's characteristic radius by a certain factor. Smaller discs will result in increased stellar accretion rates for a given disc mass while also reducing the occurrence of non-accreting discs (the ones at the bottom of the plot). In addition, smaller discs will also tend to have greater disc lifetimes because they are less affected by photoevaporation. Discs that follow the relationship of \citet{2020ApJTobinA} are found to have overly low accretion rates for their masses, while when reduced by a factor two, stellar accretion rates are now too high. The best-fit parameter is therefore between these two.

Then, the third row shows the effect of the $\alpha$ viscosity parameter. The synthetic populations shown here show less spread due to the use of a fixed value of $\alpha$, whereas the ones above have the individual values selected from a distribution that goes from \num{e-3.5} to \num{e-3.0}. The results are in line with the discussion in Sect.~\ref{sec:pop-canonical}: low values of $\alpha$ result in low accretion rates with a large amount of remaining discs, while large values result in a small number of discs with excessive stellar accretion rates. Therefore, while large values of $\alpha$ could be used to have large accretion rates, they would also require a corresponding increase in the initial disc masses to maintain the expected lifetime distribution. We find that values around $\alpha=\num{1e-3}$ provide a reasonable match to disc lifetimes and stellar accretion rates. In addition, we also note that there is an increase in the observed dust masses along with $\alpha$, which is related to the extent of the solid disc. With large $\alpha$, the grains do not grow as much as with lower $\alpha$, and as a consequence do not drift as fast. Hence, the dust emits from a larger area, which is then reflected in the emitted flux and the observed mass. However, this effect only  lasts until the dispersal of the gas disc.

\end{appendix}
\end{document}